\documentclass[iop,numberedappendix]{emulateapj}

\def\msun{\hbox{{\it M}$_{\odot}$}}
\def\mdot{\hbox{$\dot {\it M}$}}
\def\micron{$\mu$m}
\def\microns{$\mu$m}

\newcommand\be{\begin{equation}}
\newcommand\en{\end{equation}}

\newcounter{column_number}
\begin{document}

\shortauthors{Espaillat et al.}
\shorttitle{MIR Neon Emission from Disks}

\title{Tracing High-Energy Radiation from T Tauri Stars Using Mid-Infrared Neon Emission from Disks}

\author{
C. Espaillat\altaffilmark{1,2,3},
L. Ingleby\altaffilmark{4},
E. Furlan\altaffilmark{5,6},
M. McClure\altaffilmark{4},
A. Spatzier\altaffilmark{7}, 
J. Nieusma\altaffilmark{4},
N. Calvet\altaffilmark{4},
E. Bergin\altaffilmark{4}
L. Hartmann\altaffilmark{4},
J. M. Miller\altaffilmark{4},
\& J. Muzerolle\altaffilmark{8}
}

\altaffiltext{1}{NSF Astronomy \& Astrophysics  Postdoctoral Fellow}
\altaffiltext{2}{NASA Sagan Fellow}
\altaffiltext{3}{Harvard-Smithonian Center for Astrophysics, 60 Garden
Street, MS-78, Cambridge, MA, 02138, USA; cespaillat@cfa.harvard.edu}
\altaffiltext{4}{Department of
Astronomy, University of Michigan, 830 Dennison Building, 500 Church
Street, Ann Arbor, MI 48109, USA; lingleby@umich.edu, melisma@umich.edu,
jdnieusma@gmail.com, ncalvet@umich.edu, ebergin@umich.edu, lhartm@umich.edu, jonmm@umich.edu}
\altaffiltext{5}{National Optical Astronomy Observatory, 950 N. Cherry Ave., Tucson,
 AZ, 85719, USA} 
\altaffiltext{6}{Visitor at the Infrared Processing and Analysis Center, Caltech,
 770 S. Wilson Ave., Pasadena, CA, 91125, USA, furlan@ipac.caltech.edu}
\altaffiltext{7}{Oberlin College, Wright Laboratory of Physics, 
110 N. Professor St., 
Oberlin, OH, 44074, USA; aspatzie@oberlin.edu} 
\altaffiltext{8}{Space Telescope Institute, 3700 San Martin Drive,
Baltimore, MD 21218, USA; muzerol@stsci.edu}

\begin{abstract} %250 word limit

High-energy radiation from T Tauri stars (TTS) influences the amount and
longevity of gas in disks, thereby playing a crucial role in the
creation of gas giant planets. Here we probe the high-energy ionizing
radiation from TTS using high-resolution mid-infrared (MIR) {\it Spitzer} IRS Neon
forbidden line detections in a sample of disks from IC~348, NGC~2068, and Chamaeleon.  
We report three new detections of [\ion{Ne}{3}] from CS~Cha, SZ~Cha, and T~54, doubling
the known number of [\ion{Ne}{3}] detections from TTS.
Using [\ion{Ne}{3}]-to-[\ion{Ne}{2}] ratios in conjunction with X-ray
emission measurements, we probe high-energy 
 radiation from TTS.  
The majority of previously inferred [\ion{Ne}{3}]$/$[\ion{Ne}{2}] ratios based on [\ion{Ne}{3}] 
line upper limits are significantly less than 1, pointing to the dominance of either X-ray radiation
or soft Extreme-Ultraviolet (EUV) radiation in producing these lines.
Here we report the first
observational evidence for hard EUV dominated Ne forbidden line production in a T Tauri disk: 
[\ion{Ne}{3}]$/$[\ion{Ne}{2}]$\sim$1 in SZ~Cha.
Our results provide a unique insight into the EUV emission from TTS, 
by suggesting that EUV radiation may dominate the creation of Ne forbidden lines, albeit in a minority of cases.

\end{abstract}


\section{Introduction} \label{intro}

We can develop timescales for the evolution of planetary systems and
refine theories of planet formation by studying the gas in the inner,
planet-forming regions of disks around young TTS. On the one hand, significant gas must
be present for gas giant planets to form.  On the other, the lifetime of the gas in
the inner disk places an upper limit on the timescale for giant planet
formation.  Short dissipation timescales ($\sim$1000 yr) favor the
formation of planets via gravitational instability \citep[e.g.,][]{boss97} while
longer dissipation timescales allow for core accretion, which takes a
few Myr \citep[e.g.,][]{lissauer07}.  The amount and longevity of gas
in the disk is linked to the rate at which gas is eroded by
photoevaporative winds created by high energy radiation fields from the
central star \citep{hollenbach94, clarke01,alexander09}.

Initial models of photoevaporation by EUV (13.6 eV $<$ h$\nu$ $<$ 100
eV) emission predicted low mass loss rates of $\sim10^{-10}$  $\msun$
yr$^{-1}$ \citep[e.g.,][]{hollenbach94, clarke01, alexander06}. Recent models of photoevaporation
which include X-ray (0.1-2 keV) and far-ultraviolet (FUV; 6 eV $<$ h$\nu$ $<$ 13.6
eV) emission achieve mass loss rates up to
$10^{-8}\;\msun$ yr$^{-1}$ \citep{ercolano09, gorti09a, owen11}.
Observations of objects with rates much lower than this limit and still surrounded by
substantial disks challenge these recent photoevaporation models \citep{ingleby11}.  These
observations are in better agreement with the low mass loss rates
predicted by the EUV photoevaporation models. However, the main
weakness of photoevaporation models is the uncertainty in the
ionizing flux, given that no robust measurement of EUV emission from TTS currently exists.  

In TTS, most EUV emission is produced in the stellar corona and the transition region between the corona and chromosphere in active stars \citep{brown10}.  
Accreting TTS may have additional contribution to the EUV emission from the accretion shock which, in the near-UV (NUV), is significantly in excess of the UV emission from an active chromosphere for typical TTS \citep{calvet98}.
Our knowledge of the strength of EUV emission from TTS is limited due to absorption of EUV photons by neutral hydrogen along
the line of light.  
In accreting TTS, EUV emission may be absorbed in the accretion flows, which have high column densities ($n_H$) up to $10^{22}$ cm$^{-3}$; even for non-accreting TTS, the interstellar hydrogen column density dampens the transmission of EUV photons.  For example, the column density of neutral hydrogen toward TW Hya, the closest accreting TTS, is $\sim$5$\times$10$^{20}$ cm$^{-3}$ \citep{kalberla05,dickey90}.   According to \citet{drake99}, with these column densities, less than 10\% of the EUV flux at 100 {\AA} will be transmitted through the interstellar medium, making it difficult to ascertain the EUV spectrum.

A mission devoted to EUV observations, the Extreme Ultraviolet Explorer (EUVE) All Sky Survey was unsuccessful in detecting TTS. However, it did obtain spectra of many active stars, including RS CVn stars \citep{sanzforcada12} and dwarf stars which undergo strong flares \citep{brown10}.  The EUV spectra of these active stars are sometimes used when approximating the radiation field incident on circumstellar disks \citep{owen10}.  Whether or not this is an appropriate proxy for a TTS's EUV radiation field is unclear.

As a result, secondary methods for estimating the strength of the EUV
radiation field have been sought. \citet{alexander05} used FUV
He II $\lambda$1640 {\AA} line fluxes to probe the strength of
ionizing emission at $<$228 {\AA}.  
However, Ardila et al. (submitted) argue that
other mechanisms besides radiative
recombination may play a role in the excitation of the line.
For example, \citep{brown81}
find that collisional excitation plus photon trapping
were the dominant agents in He II $\lambda$1640 {\AA}
formation in the case of T~Tau. Another alternative
is to use observations of solar type
stars at different ages to infer the evolution of high energy radiation
fields.
\citet{ribas05} analyzed high
energy emission from the X-ray to the FUV of nearby young stars
between 0.1 and 7 Gyr, from which they developed power law fits to the
decay in high energy radiation fields.  Due to the proximity of the
targets, they were able to use EUV spectra obtained with EUVE, but
even then, most of the detected flux was in emission lines.  The
strength of the emission lines (mainly highly ionized Fe lines) was
observed to decrease with age.  It is unclear whether the
\citet{ribas05} power laws for EUV radiation may be extended down to the
ages of T Tauri stars (1--10~Myr).  Given that X-ray and FUV emission likely
saturate at ages $<$10--30 Myr \citep{preibisch05,ingleby12}, 
EUV emission may saturate as well.

An alternative high-energy radiation diagnostic line emerged with the detection of
mid-infrared (MIR) fine-structure forbidden emission lines of \ion{Ne}{2} with the Infrared
Spectrograph \citep[IRS;][]{houck04} on the {\it Spitzer Space Telescope} \citep{werner04} in over 50 TTS
\citep{pascucci07, espaillat07a, lahuis07, flaccomio09, gudel10,
baldovin11,szulagyi12}. 
Due to the high first ionization potential of Ne (21.56~eV), [\ion{Ne}{2}]
is capable of constraining high energy stellar radiation.
Two mechanisms have been proposed to be responsible for [\ion{Ne}{2}]
lines in disks around TTS; while some researchers have posited that MIR Ne fine-structure
lines can be attributed solely to X-ray ionization and heating
\citep[][GNI07]{glassgold07}, recent work by \citet[][HG09]{hollenbach09} has shown that EUV photons can play an
important role in creating [\ion{Ne}{2}] lines. 

In the GNI07 thermal-chemical model, the gas in the upper atmosphere of the
protoplanetary disk is heated and ionized by stellar X-rays. 
GNI07 and \citet{meijerink08} have shown that X-rays can heat the gas
in the disk to temperatures up to $\sim$4000-5000~K at an altitude where N$_{H}$$\sim$10$^{21}$ cm$^{-2}$
 out to about 25~AU from the star.  These authors find that the degree of ionization in these regions is less than a few percent.
These two works
consider only X-rays for ionizing the neutral and low-ionization forms
of the Ne atom since FUV photons have insufficient energy to ionize
Ne and EUV is absorbed over short distances. 
More recently, 
HG09 argue that
EUV heating should not be discounted given that it can penetrate disk
winds when the mass accretion rate onto the star is less than 10$^{-8}$
${\msun}$ yr$^{-1}$, the average accretion rate in TTS
\citep{hartmann98}. 
HG09 have shown that EUV photons can potentially play an
important role in creating [\ion{Ne}{2}] lines as well.  These lines can originate in
the EUV surface layer of the disk which is completely ionized and reaches temperatures as high as 10,000~K \citep[e.g.,]{hollenbach94}.

Since the first detections of MIR Ne fine-structure lines, 
researchers have attempted to link
observed [\ion{Ne}{2}] line luminosities to
high-energy stellar radiation.  
\citet{pascucci07} found a correlation between the [\ion{Ne}{2}] luminosity and
the X-ray luminosity, supporting X-rays as the dominant [\ion{Ne}{2}] production
mechanism.  However, \citet{espaillat07a} 
did not find such a trend between [\ion{Ne}{2}] luminosities and X-ray luminosities.  Instead, they found a correlation
between the [\ion{Ne}{2}] luminosity and the accretion rate.
\citet{espaillat07a} proposed that this was evidence that EUV radiation
was important in creating [\ion{Ne}{2}]  lines given that  the accretion shock on the stellar
surface is expected to create EUV emission, as mentioned earlier.
Moreover, \citet{calvet04} found that FUV fluxes scale 
with L$_{acc}$ and a similar scaling is expected for EUV emission.
Later work by \citet{gudel10} with a much larger sample,
found correlations between the [\ion{Ne}{2}] luminosity and
both the X-ray luminosity and accretion rate.  However, these correlations have large scatter, possibly due to
different disk structures \citep{schisano10} or variability.  
Based on known [\ion{Ne}{2}] detections, it is difficult to distinguish 
between the relative importance of X-ray and EUV radiation in creating Ne fine-structure line emission in disks without additional data.

Here we approach the issue of linking high-energy TTS radiation to disk gas
emission from a different avenue with new [\ion{Ne}{3}] emission line
detections.  Ne has a second ionization potential of 41.0~eV, and [\ion{Ne}{3}], like [\ion{Ne}{2}], can
originate in either the X-ray or EUV layer of the disk (GNI07, HG09).
Due to the limited number of [\ion{Ne}{3}]
detections in the past \citep{lahuis07, flaccomio09, najita10}, the
potential of this line to trace high energy radiation has not been significantly utilized.  
Recently, \citet{szulagyi12} made an extensive search for [\ion{Ne}{3}] in 56 disks around TTS and found no detections.
This highlights the rarity of this line and the importance of using new [\ion{Ne}{3}] line detections to constrain 
the radiation field irradiating the circumstellar disk.
In Section~2, we detail our sample selection and review the data
presented in this work.  In Section~3, we present three new [\ion{Ne}{3}] emission line detections in our sample (Section~3.1) and search for
correlations between X-ray radiation and [\ion{Ne}{3}] emission lines (Section~3.2). 
We discuss the implications of our results on the connection between
high-energy TTS radiation and the disk in further detail in Section~4.

\section{Observations \& Data Reduction} \label{redux}

\subsection{Sample Selection} \label{sampleselection}

Our sample consists of 19 objects in NGC~2068, IC~348, and
Chamaeleon (Table~\ref{tab:targetlog}). 
Chamaeleon is a $\sim$2~Myr old
region with relatively isolated star-formation and low
extinction \citep{luhman08}. NGC~2068 and IC~348 are $\sim$2~Myr old
\citep{flaherty08} and $\sim$3~Myr old \citep{luhman03}, respectively,
and are more clustered star-forming regions with higher extinction. The
main goal of our project was to detect the [\ion{Ne}{2}] line using
high-resolution {\it Spitzer} IRS spectra.  Therefore, we selected disks
where it would be easier to detect line emission, namely those that
exhibited low continuum emission such as transitional disks (TD), 
pre-transitional disks (PTD), and full disks with
low continuum emission \citep{espaillat12a}.

We identified the disks for this sample using the criteria above and
existing low-resolution IRS spectra. This sample contains 6 FD, 6 TD,
and 7 PTD (Table~\ref{tab:prop}), all of which have known disk properties determined by
SED fitting using the \citet{dalessio06}
irradiated, accretion disk model. The low-resolution IRS
spectra for our NGC~2068 and IC~348 objects were presented in
\citet{espaillat12a} along with detailed SED modeling.  For our Chamaeleon targets,
low-resolution IRS spectra were presented in \citet{kim09} and
\citet{manoj11}.  In \citet{espaillat11}, we presented detailed modeling
of the SEDs of CS~Cha, SZ~Cha, T~35.  Modeling of T~54 was presented in
\citet{kim09}.  Stellar parameters for our targets and references
for these quantities are listed in Table~\ref{tab:prop}.

\subsection{Data Reduction}

Here we present high-resolution {\it Spitzer} spectra for our
19 targets as well as X-ray data for
most of our sources in IC~348 and NGC~2068.  

\subsubsection{Infrared Data}

{\it Spitzer} IRS spectra for all of our targets except CS~Cha 
were taken in General Observing (GO) Program 40247 (PI: Calvet; Table~\ref{tab:irslog}).
We also included archival IRS observations of CS~Cha
from Program 30300 (PI: Najita; Table~\ref{tab:irslog}). 
All of the observations were performed in staring mode using the
short-high-resolution module (SH) of IRS, spanning wavelengths from
10--19~{\microns}, with a resolution $\lambda/\delta\lambda\sim$600.
                                                                                                                       
Details on the observational techniques and general data reduction
steps, including bad pixel identification, sky subtraction, and flux
calibration, can be found in \citet{furlan06} and \citet{watson09}.
Here we provide a summary. Each object was observed twice along the
slit, at a third of the slit length from the top and bottom edges of the
slit. Instead of using the basic calibrated data (bcd) products from
the {\it Spitzer} Science Center pipeline, we used the droop products,
pipeline version S18.18. As opposed to the bcd products, the droop
products lack corrections for inter-order light leakage (which is a 
very small effect) and are not divided by the flatfield. We found that 
especially the latter step introduced more noise to our data, so the 
droop products yielded spectra with better signal-to-noise ratios. 

We reduced the data with the SMART package \citep{higdon04}
and used the ISO Spectral Analysis Package (ISAP) within SMART
to measure line fluxes, uncertainties, and upper limits. 
Bad and rogue pixels were corrected by interpolating
from neighboring pixels. 
For each of the targets, we obtained several observations at each nod 
position to increase the time spent on the target. We also acquired an 
off-source background observation for each target.
We extracted each background-subtracted spectrum individually, then calculated
their average and standard deviation.
After background subtraction, a full slit extraction was
performed. To flux calibrate the observations we used spectra of $\xi$
Dra (K2III).  We performed a nod-by-nod division of the target spectra
and $\xi$ Dra spectra and then multiplied the result by a template
spectrum \citep{cohen03}. The final spectrum was produced by averaging
the calibrated spectra from the two nods, and uncertainties were derived
from the standard deviation of the mean flux at each wavelength.
Our spectrophotometric accuracy is typically 2--5$\%$. We note that we 
manually masked artifacts in the extracted, calibrated spectra that 
were not captured by the bad pixel and rogue masks. We also applied
extinction corrections to the IRS data using the A$_{V}$ for each object
listed in Table~\ref{tab:prop}.

The average signal-to-noise ratio (SNR) in the continuum for our spectra
is 30, with a range of 15--70.  We note that FM~281 has
SNR$\sim$5 and so we exclude these data
from our analysis.  We overlaid the IRS slit positions for each AOR on K-band images
from the Two Micron All Sky Survey \citep[2MASS;][]{skrutskie06}
to check for anomalous behavior. In the observation of LRLL~21 in IC
348, another object was located in one of the nods. However, our SH
spectrum (with a slit width of 4.7$^\prime$$^\prime$) agrees with an
earlier GTO SL spectrum (slit width = 3.6$^\prime$$^\prime$; AOR ID:
16269056) which does not include this object.  We conclude that this
other object that entered the slit does not emit significantly in the
mid-infrared and so object LRLL~21 dominates the SH spectrum presented
here.

\subsubsection{X-ray Data} \label{xprop}

Of the 19 targets in our  sample, X-ray properties for 18 of
them were either extracted in this work from {\it Chandra} observations
or compiled from the literature. 
X-ray luminosities for the Chamaeleon targets
were taken from the literature (Table~\ref{tab:prop}).   X-ray properties for objects in IC~348 and 
NGC~2068 come from this work. {\it Chandra}
observations were obtained as part
of a joint Chandra-Spitzer GO program (Proposal 09200909, PI: Calvet)
for IC 348 (Observation ID (Obs ID) 8584) and NGC 2068 (Obs ID 8585 and 10763).
We note that our {\it Spitzer} and {\it Chandra}
data were not taken simultaneously.

The above observations were performed with the {\it Chandra} Advanced
CCD Imaging Spectrometer (ACIS) in FAINT mode using the Imaging Array
(ACIS-I) and Spectroscopic Array (ACIS-S). Dates and exposure times can
be found in Table~\ref{tab:chandralog}. The {\it evt2} files analyzed
here were obtained through the {\it Chandra} processing pipeline
(version: 7.6.11.4). Events and spectra for targets in our 
sample were extracted using the
\anchor{http://www.astro.psu.edu/xray/acis/acis_analysis.html}{{\em ACIS
Extract}} (AE) software package (version 2010-02-26)\footnote{ The {\em
ACIS Extract} software package and User's Guide are available at
\url{http://www.astro.psu.edu/xray/acis/acis$\_$analysis.html}} and the
{\it ae$\_$better$\_$backgrounds} algorithm. \citep{broos10}.

We measured X-ray fluxes for our NGC~2068 and IC~348 targets by
extracting X-ray spectra using the HEASARC X-ray fitting package XSPEC 
version 12.6 \citep{arnaud96} and AE fitting scripts. Source counts over
the energy range 0.5--8 keV are listed in
Table~\ref{tbl:thermal_spectroscopy}.  Most of the sources have more
than 100 counts, the exceptions being LRLL~72, LRLL~133, and FM~281.
LRLL~2 and LRLL~6 have over 1,000 counts.

We fit all of the spectra with single- and two-temperature {\it vapec}
thermal collisional ionization equilibrium plasma models \citep{smith01}
along with an absorbing column of interstellar material  \citep[i.e.
{\it tbabs} absorption model;][]{wilms00}. Automated fitting performed by
AE adopted elemental abundances frozen at the values used by the XEST
study for typical pre-main sequence stars \citep{gudel07}. We left the
temperature(s) and the absorbing hydrogen column density, $N_H$, as free
parameters. In most cases, the $\chi^2$ between the one- and
two-temperature fits were similar.  In
Table~\ref{tbl:thermal_spectroscopy} we list the parameters from the
best fitting one-temperature model.  However, for LRLL~67, FM~177, and
FM~618, the $\chi^2$ of the two-temperature model was 2--3 times better
than that of the one-temperature model.   For these three objects we
list the parameters of the best fit two-temperature model in
Table~\ref{tbl:thermal_spectroscopy}. The luminosities were derived from
spectral modeling assuming a distance of 315~pc for IC~348  \citep{luhman08} and 400~pc
for NGC~2068 \citep{flaherty08}.

We note that FM~515 was not within the FOV of the observations and was
not located in the {\it Chandra} or {\it XMM-Newton} data archive. 
Therefore, we do not provide an X-ray flux for FM~515 here. LRLL~133 was
located within the FOV of our IC~348 observation, but was not detected
while LRLL~21 was not within the FOV.  However, LRLL~133 and LRLL~21
were detected in ObsID 606 (53~ks; 2000-09-21; PI: Preibisch) and these
data were used to extract the X-ray fluxes for these two objects in
Table~\ref{tbl:thermal_spectroscopy}. LRLL~68 was within the FOV of our
observations, but was not detected and was not observed by another
program.  We estimate an upper limit for the luminosity of this source
by calculating the on-axis limiting sensitivity of ObsID: 8584 with the
Portable Interative Multi-Mission Simulator (PIMMS)\footnote{
\url{http://heasarc.gsfc.nasa.gove/Tools/w3pimms.html}}.  We assume an
on-axis detection of 3~counts over the energy range 0.5--8~keV in our
50.14~ks exposure and a thermal plasma of 2~keV.  We adopt an absorbing
column of 1.6$\times$10$^{22}$~cm$^{-2}$, calculated with the $N_H$ and
$A_J$ relation of \citet{vuong03}, along with an A$_J$ of 0.83 from
\citet{luhman03}.  The derived limiting flux is
7.4$\times$10$^{-16}$~ergs~s$^{-1}$~cm$^{-2}$ and the limiting
absorption-corrected flux is
1.9$\times$10$^{-15}$~ergs~s$^{-1}$~cm$^{-2}$. We adopt an upper limit
of 2.3$\times$10$^{28}$~ergs~s$^{-1}$ for the absorption-corrected X-ray
luminosity of LRLL~68.

\section{Results} \label{results}

Here we search for Neon forbidden emission in our sample.
We explore the relationship between [\ion{Ne}{3}] lines and high-energy
radiation by looking at the X-ray properties of TTS with [\ion{Ne}{2}].  In
our sample of disks with [\ion{Ne}{3}], we
also compare X-ray properties to observed [\ion{Ne}{3}]-to-[\ion{Ne}{2}] ratios.

\subsection{Neon Line Detections} \label{detected}

\begin{figure}
\epsscale{1}
\plotone{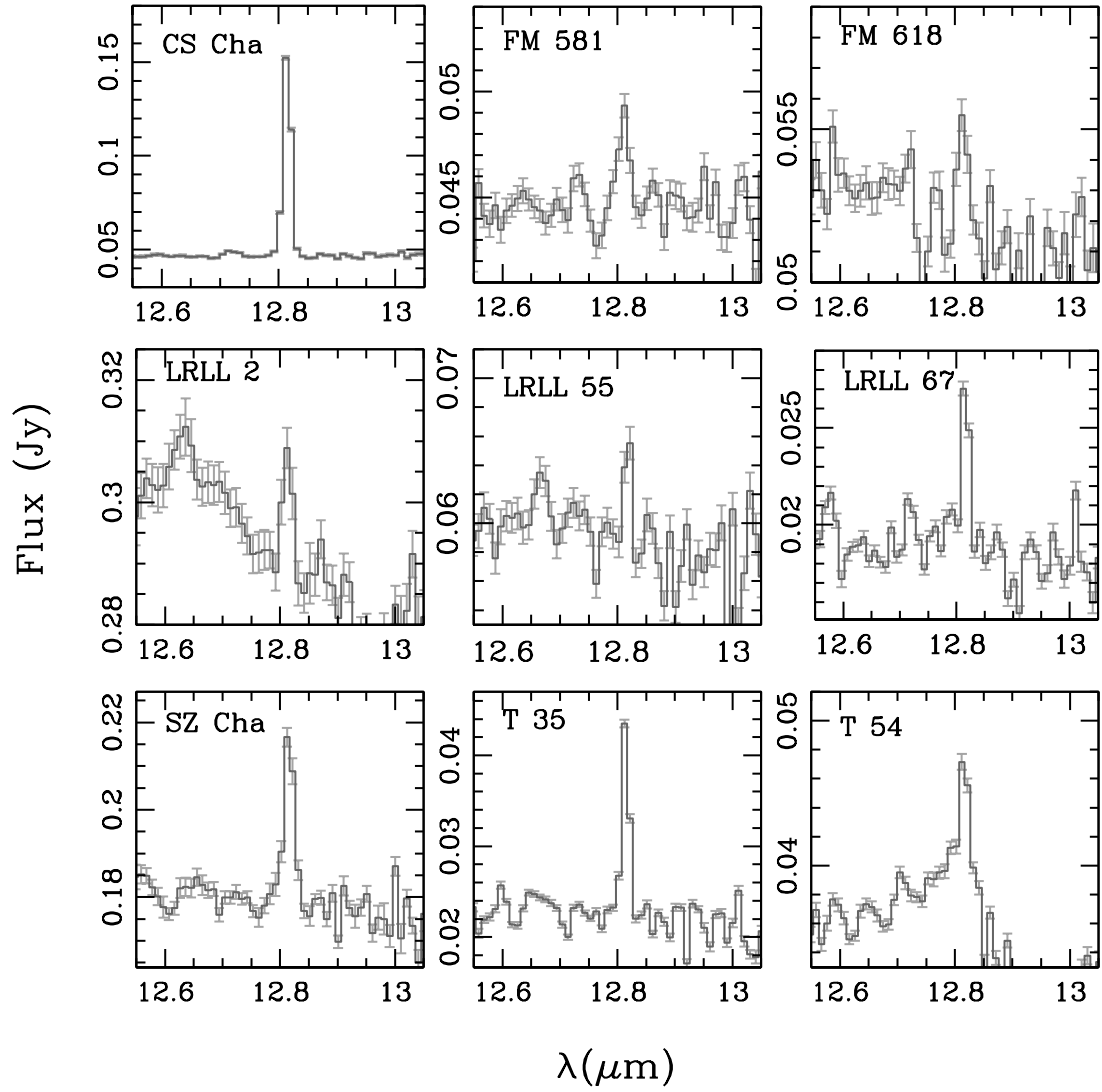}
\caption[]{
[Ne~II] detections at 12.81~{\micron} in our {\it Spitzer} SH spectra.  
In LRLL~2,  the residual PAH emission at 12.7~{\micron} is due to imperfect background subtraction.   
}
\label{figneii}
\end{figure}

\begin{figure}
\epsscale{1}
\plotone{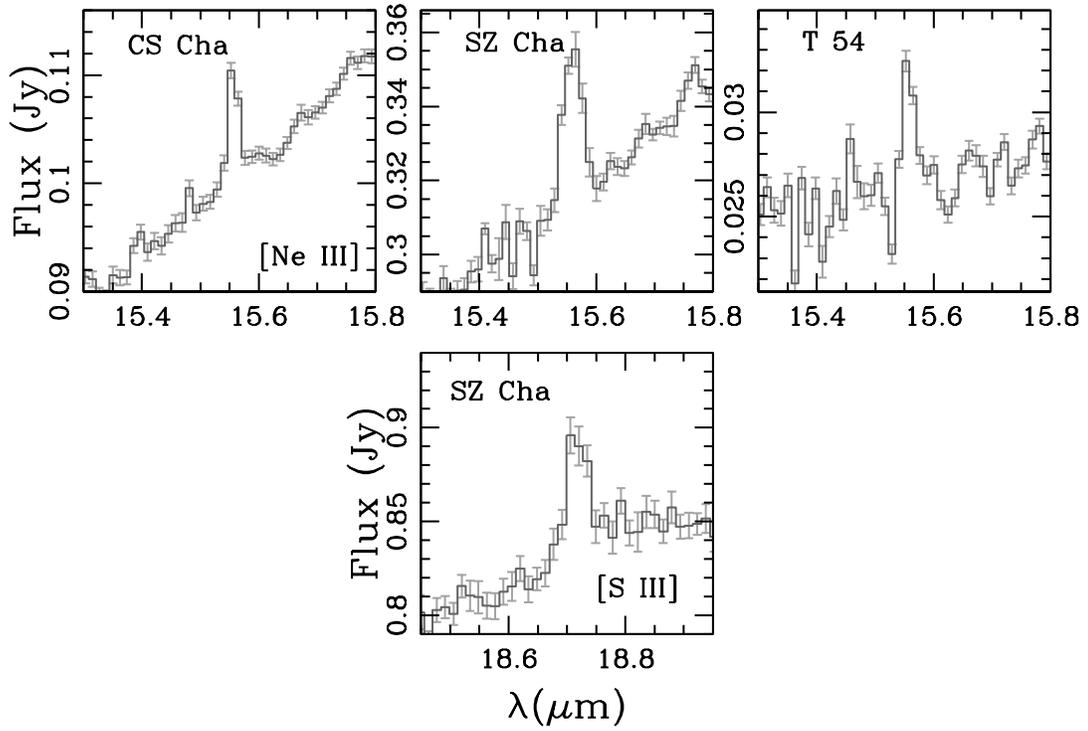}
\caption[]{
[Ne~III] and [S~III]  detections in our {\it Spitzer} SH spectra at 15.55~{\micron} and 18.71~{\micron}, respectively.   
}
\label{figneiiisiii}
\end{figure}

We detected [\ion{Ne}{2}] line emission from nine objects in our 
sample (Figure~\ref{figneii}).  Line fluxes, errors, and SNR are listed in
Table~\ref{tab:neon} along with 3$\sigma$ upper limits for sources with
non-detections. 
[\ion{Ne}{3}] lines are also present in three objects in our sample: CS~Cha, SZ~Cha,
and T~54 (Figure~\ref{figneiiisiii}; Table~\ref{tab:neon}).  
[\ion{S}{3}] is also detected in SZ~Cha with a line flux of 2.63$\pm$0.3$\times$10$^{-14}$
erg~cm$^{-2}$~s$^{-1}$ and SNR$\sim$12 (Figure~\ref{figneiiisiii}).
We note that the 13$_{6~8}$ to 12$_{3~9}$ transition of H$_{2}$O is located at $\sim$15.57~{\microns}.  However,
neither SZ~Cha, CS~Cha, nor T~54 have evidence for any other water lines in their spectra.  It is
highly unlikely that only one of the numerous H$_{2}$O line transitions would
be present  \citep[see Figure~4 of][]{pontoppidan10}.

In SZ~Cha the [\ion{Ne}{3}] and  [\ion{S}{3}]  lines are broader than 
the [\ion{Ne}{2}] line.  The FWHM of [\ion{Ne}{3}] and [\ion{S}{3}] is $\sim$700 km s$^{-1}$.  The 
FWHM of [\ion{Ne}{2}]  is $\sim$500 km s$^{-1}$, essentially equal to the resolution
of the IRS SH module.
Inspection of the data shows that these lines are not spatially extended.
However, one pixel is 2.3$^\prime$$^\prime$ or about 370 AU at 160 pc.
Therefore, we cannot exclude the possibility
that these broad line widths indicate more than one emitting region within
$\sim$370~AU (e.g. disk, wind).
\citet{najita09} suggested the [\ion{Ne}{2}] line flux was composed of both a broad,
extended shock component and a spatially unresolved disk component
based on ground-based high-resolution spectra of two different disks.
Likewise, higher spectral and spatial resolution data of SZ~Cha are needed to explore
this further.

\subsection{[\ion{Ne}{3}] \& High-Energy Radiation} \label{highe}

Compared to [\ion{Ne}{2}], [\ion{Ne}{3}] line emission is much
less common.  Prior to this work, there were only three reports of
[\ion{Ne}{3}] lines: Sz 102 \citep{lahuis07}, WL~5
\citep{flaccomio09}, and TW Hya \citep{najita10}. Here we reported three
new detections in SZ~Cha, T~54, and CS~Cha.  
We limit
ourselves to objects with disks in this work and exclude the Class~III
object WL~5\footnote{ \citet{mcclure10} classified WL~5 as a Class~III
object based on its low-resolution IRS spectrum. The 
spectrum of WL~5 presented in \citet{mcclure10} is a composite of two
objects.  In the short-low-resolution module of IRS (SL;
5--14~{\micron}), WL~5A was the only object in the slit while in the
long-low-resolution module of IRS (LL;14--38~{\micron}) its companion
(8.5$^\prime$$^\prime$ away) entered the slit. WL~5A is a photosphere
based on the SED slope in the SL data. The companion is most likely a
Class~I object based on the steeply rising SED observed in the LL data.
Given the orientation of the SH slit used in \citet{flaccomio09}'s
observation of WL~5A and the fact that a nearby background observation 
was not taken for data reduction, it is likely that some emission from
the nearby Class~I object entered the slit.  Therefore, the Ne fine-structure line
emission from WL~5 is most likely from the
nearby Class~I object, especially given that Class~III objects are typically not surrounded by
material.}.  To the best of our knowledge, there
are no studies which search for jets in NGC 2068 and IC 348.  
SZ~102 and CS~Cha are known
to have jets \citep[][]{lahuis07,takami03,gudel10}.

Since MIR Ne fine-structure lines have been proposed to be due to X-rays (GNI07,
HG09), we first tested if sources with [\ion{Ne}{3}] have a higher X-ray
luminosity than sources that have [\ion{Ne}{2}] but no
[\ion{Ne}{3}] detections. Our
sample in Figure~\ref{figxray} is composed of objects in this work and
the literature which have reported [\ion{Ne}{2}] detections and X-ray
luminosities \citep[see the Appendix for more details;][]{gudel10, baldovin11, carr11, sacco12, lahuis07, najita10, pascucci07, ingleby11, white00, neuhaeuser95}. 
We find that the X-ray luminosities of objects with [\ion{Ne}{3}] lines
is similar to those of objects with only [\ion{Ne}{2}] lines. 
The median X-ray luminosity of
sources with [\ion{Ne}{2}] but without [\ion{Ne}{3}]  is 1.6$\times$10$^{30}$ ergs~s$^{-1}$.  For those
with [\ion{Ne}{3}] lines, the median is 2.1$\times$10$^{30}$
ergs~s$^{-1}$.  Given that the KS probability these are both taken from the
same distribution is 99$\%$, we conclude that sources with
[\ion{Ne}{3}] emission are not significantly more luminous in X-rays than
sources without [\ion{Ne}{3}] emission.   However, we note that given the
small number of known sources with  [\ion{Ne}{3}] emission, this needs to 
be confirmed with a larger sample.

[\ion{Ne}{2}]  and [\ion{Ne}{3}]  can be also be due to EUV radiation from the
central star (HG09). One way to distinguish between X-ray and EUV
creation of Ne fine-structure emission is by measuring the [\ion{Ne}{3}]-to-[\ion{Ne}{2}]
line ratio.  Ne ion production by  X-ray emission can lead to [\ion{Ne}{3}]-to-[\ion{Ne}{2}] line ratios of
the order $\sim$0.1 \citep[see Fig. 4 of GNI07 and Fig. 17][]{meijerink08}.  However, EUV photons can lead to much
higher [\ion{Ne}{3}]-to-[\ion{Ne}{2}] line ratios (HG09).
This is largely because X-ray photons penetrate
deeper into the disk than EUV photons.  The GNI07 and HG09 models find that the X-ray dominated layer of the disk
is predominantly neutral.  The abundance of H atoms leads to efficient
charge exchange of Ne$^{++}$ and H, leading to Ne$^{+}$.
Therefore, in the X-ray layer of the disk, there will be more \ion{Ne}{2} than \ion{Ne}{3}, leading
to larger [\ion{Ne}{2}] line luminosities relative to those of [\ion{Ne}{3}].
In contrast, the models find that the EUV regions are fully ionized; the abundance of H atoms is low and so here 
charge exchange is not that efficient.
Therefore, in the EUV layer of the disk there will be more \ion{Ne}{3} relative to the X-ray layer and
hence it is possible that the [\ion{Ne}{3}] line luminosity can be greater than the [\ion{Ne}{2}] line luminosity.

Prior to our work, all objects with [\ion{Ne}{3}] emission had [\ion{Ne}{3}]$/$[\ion{Ne}{2}]$<$0.1,
in line with X-ray dominated Ne ion production.
Here we report the first source with
[\ion{Ne}{3}]$/$[\ion{Ne}{2}]$\sim$1, SZ~Cha, which has a ratio of 1.36$\pm$0.24.  
We note that T~54 also has [\ion{Ne}{3}]$/$[\ion{Ne}{2}]$>$0.1.  However,
T~54's ratio of 0.34$\pm$0.13 is not 3$\sigma$ above 0.1 within the uncertainties
of the line measurement; future observations with higher sensitivity are needed to confirm this ratio.
The value of [\ion{Ne}{3}]$/$[\ion{Ne}{2}] can also trace the shape of the EUV spectrum.
[\ion{Ne}{3}]$/$[\ion{Ne}{2}]$\sim$1 points to roughly L$_{EUV}$$\sim$$\nu^{-2}$  (Fig 1. of HG09; U. Gorti, personal
communication).  A caveat is that it is likely X-rays are present and will contribute to the observed [Ne~II]
line emission, contaminating the EUV spectral slope determination. 

Given that the EUV luminosity has to be
two times greater than the X-ray luminosity to dominate Ne ion production
(HG09), we compared the [\ion{Ne}{3}]-to-[\ion{Ne}{2}] line ratio to the X-ray
luminosity of the object, normalized by its bolometric luminosity (L$_{bol}$; Figure~\ref{figreln}).  
We note that here we use L$_{bol}$ to refer to the stellar luminosity (L$_{*}$). 
L$_{X}$ and L$_{*}$ for SZ~Cha, CS~Cha, and T~54 are listed in Table~\ref{tab:prop}.
TW Hya's L$_{X}$ and L$_{*}$ are from \citet{ingleby11}.  SZ~102's
L$_{X}$ is from \citet{gudel10}; we estimated L$_{*}$ using
2MASS photometry and \citet{kh95} colors for an M0 star \citep{lahuis07}.
We note that CS~Cha and T~54 are known binaries. CS~Cha is a spectroscopic binary with a separation of $\sim$4~AU \citep{guenther07} and both components are
of equal brightness \citep{nguyen12}.  Neither component was resolved in the optical nor X-ray.
T~54 is brighter than its secondary by a factor of at least 5 and their separation is $\sim$40~AU \citep{lafreniere08}.
Using $L_{bol}$ based on the primary star should not significantly affect the correlation we find here, especially given that 
we assume that the error in L$_{X}$$/$$L_{bol}$
is dominated by the variability in the X-ray.  We adopt
an error of a factor of 2 for L$_{X}$ based on the typical range in X-ray variability, as has been done in previous works \citep[e.g.,][]{gudel10}.

In Figure~\ref{figreln}, [\ion{Ne}{3}]$/$[\ion{Ne}{2}]
increases as L$_{X}$$/$L$_{bol}$ decreases.  We find a strong trend (with a Pearson correlation coefficient of -0.9)
and fit the data with the following equation:
\begin{equation}
log\frac{[Ne~III]}{[Ne~II]} = (-0.98\pm0.29)log\frac{L_X}{L_{bol}}+(-3.78\pm0.89).
\end{equation}
This trend is consistent with EUV radiation
dominating the production of [\ion{Ne}{2}] and [\ion{Ne}{3}]  at low L$_{X}$$/$L$_{bol}$ values in Figure~\ref{figreln}  and leading to a high
[\ion{Ne}{3}]$/$[\ion{Ne}{2}].  It follows that X-ray radiation dominates the heating of the
disk at higher L$_{X}$$/$L$_{bol}$ values in Figure~\ref{figreln}, creating more [\ion{Ne}{2}] than
[\ion{Ne}{3}] (i.e., [\ion{Ne}{3}]$/$[\ion{Ne}{2}]$<<$1).
We note that this trend is promising, but largely driven by SZ~Cha.  Future work with archival high-resolution
IRS spectra and JWST may reveal additional objects with [\ion{Ne}{3}]$/$[\ion{Ne}{2}]$>$0.1 and known
L$_{X}$ that can be used to confirm this relation.

Another potentially important diagnostic of high-energy radiation 
could lie in the relationship between
[\ion{Ne}{3}]$/$[\ion{Ne}{2}] and [\ion{S}{3}].
Here we detected [\ion{S}{3}] 
in SZ~Cha, the disk in our sample with the highest 
[\ion{Ne}{3}]$/$[\ion{Ne}{2}].  
S$^{++}$ and Ne$^{++}$ have similar charge exchange rates with H 
\citep{butler80}.  
Therefore, for the same reasons outlined earlier, we would expect more S~III when 
EUV dominates the production of \ion{Ne}{2} and \ion{Ne}{3}.
Prior to this work [\ion{S}{3}] had only been detected in the disk
of SSTc2d$\_$J1829282 \citep{lahuis07}.
Using [\ion{Ne}{3}] upper limits from \citet{lahuis07} we estimate [\ion{Ne}{3}]$/$[\ion{Ne}{2}]$<$0.18, 
in contrast to the large [\ion{Ne}{3}]$/$[\ion{Ne}{2}] observed for SZ~Cha.  
We leave it to future work to explore any potential connection
between these lines and high-energy ionizing radiation.  

\section{Discussion} \label{sec:discuss} \label{dis:atomic}

There has not yet been a definitive observational result discriminating between X-ray and EUV radiation
as the primary source of Ne forbidden line emission. 
There have been over 50 objects with [\ion{Ne}{2}] line detections to date \citep{pascucci07, espaillat07a, lahuis07, flaccomio09, gudel10, baldovin11,szulagyi12}.
There have also been many theoretical works examining the importance of the Ne line luminosities
and their connection to high-energy radiation \citep{glassgold07,meijerink08,ercolano08, gorti08, hollenbach09, schisano10, ercolano10}.
However, comparing observations and theories has not led to clear results.  
There have been no distinct correlations between [\ion{Ne}{2}] line luminosities and L$_{X}$ or mass accretion rates.
In the case of [\ion{Ne}{2}], it is not obvious that expanding the known sample of disks
with these lines will lead to substantially clearer results.
We note that the L$_{X}$ and mass accretion rates for objects in our sample with detected [\ion{Ne}{2}] falls within range of those objects with [\ion{Ne}{2}] line detections in literature and that our results are consistent with previous results.  In addition, the models of HG09 and GNI07 use L$_{X}$$\sim$10$^{28}$--10$^{32}$ ergs s$^{-1}$ in their models, covering the range seen in the observations. 

Another relatively unexplored avenue to probe the relative importance of X-ray and EUV radiation in creating
Ne forbidden line emission
lies in [\ion{Ne}{3}] line detections.
Of those disks with [\ion{Ne}{2}] line detections, only five disks
have [\ion{Ne}{3}] line detections, three of which come from this work.  While [\ion{Ne}{3}] detections are very rare, we can start to make preliminary studies of these lines with the expectation that these results can be greatly expanded upon with {\it JWST}.   Out of the six objects with known [\ion{Ne}{3}] line emission, five of these have [\ion{Ne}{3}]$/$[\ion{Ne}{2}]$<$1.  
Also, based on [\ion{Ne}{2}] and [\ion{Ne}{3}] upper line limits, \citet{szulagyi12} find that typically [\ion{Ne}{3}]$/$[\ion{Ne}{2}]$<$1.  This ratio can either be a result of X-ray or soft EUV as the primary source of Ne forbidden lines according to the models of HG09.
We note that our report of CS~Cha's [\ion{Ne}{3}]$/$[\ion{Ne}{2}] agrees with previously observed ratios for TW~Hya and SZ~102, i.e, ratios about 0.1 or less, of the order theoretically predicted by GNI07 and \citet{meijerink08}.
T~54's ratio of 0.3 is higher than that of CS~Cha, TW~Hya, and SZ~102.  However, \citet{ercolano10} get [\ion{Ne}{3}]$/$[\ion{Ne}{2}] up to 0.3 in TD since the gas in these disks tends to be warmer than that of FD.
Studies of other lines like [\ion{Ar}{2}] have [\ion{Ar}{2}]-to-[\ion{Ne}{2}] ratios which indicate either soft EUV or soft X-ray emission as the primary source of these lines \citep{szulagyi12}.

SZ~Cha is the only known object to date with [\ion{Ne}{3}]$/$[\ion{Ne}{2}]$\sim$1;  
HG09 have shown that this high ratio can only occur in the EUV layer of the disk.
If confirmed, this result in SZ~Cha would be the only clear observational result discriminating 
between X-ray and EUV radiation as the primary source of Ne forbidden line emission. 
Some caveats to bear in mind are that the charge exchange coefficients used in the theoretical
models need to be updated \citep{butler80,ercolano10}.
In addition, high-resolution spectroscopy of the [\ion{Ne}{3}] line in SZ~Cha is necessary to constrain where 
in the disk this emission is originating from.   Lastly, a larger sample of disks with [\ion{Ne}{3}] line detections is necessary
to test this further.

While [\ion{Ne}{3}]$/$[\ion{Ne}{2}]$\sim$1 is an indication of
EUV-dominated production of Ne ions, the issue still remains that EUV
radiation is easily absorbed in the environment of the star and may not
even reach the disk.  However, EUV radiation should be able to reach the
disk in the case of SZ~Cha given its substantial accretion rate
\citep[2.4$\times$10$^{-9}$ ${\msun}$ yr$^{-1}$ as measured from its
U-band excess by][]{espaillat11}. This is because in order for the EUV
radiation to penetrate the protostellar wind, the outflow mass loss rate
must be less than 10$^{-9}$ ${\msun}$ yr$^{-1}$ (HG09). This corresponds
to a mass accretion rate of 10$^{-8}$ ${\msun}$ yr$^{-1}$ given that
protostellar winds of TTS are typically $\sim$0.1 of the mass accretion
rate \citep{hartigan95,calvet97}. Therefore, EUV radiation should reach
the disk of SZ~Cha given its accretion rate.

These Neon forbidden line observations raise the question of whether EUV
photoevaporation is responsible for the gapped disk structure of SZ~Cha.
Based on SED modeling, SZ~Cha has an inner disk separated
from the outer disk by a $\sim$20~AU gap \citep{espaillat11}. 
EUV photoevaporation
models predict that the disk wind will open a small, short-lived gap within the outer disk before clearing out an inner hole and then the
rest of the outer disk  \citep[e.g.,][]{clarke01}.  
However, the gaps from photoevaporation are expected to open quickly relative to the lifetime
of the disk, and so it is unlikely that we are catching
the disk in this short-lived phase \citep{alexander06}.  Instead, we propose that the gap in SZ~Cha was formed through dynamical
clearing by either planetary \citep{zhu11} or stellar mass
\citep{artymowicz94} companions {\it before} EUV photoevaporation took
effect.  In the case of planetary mass companions, one can speculate
that their presence led to a decrease in the mass accretion rate
\citep[e.g.,][]{lubow06,rice06} necessary for EUV radiation to reach
the disk, accelerating photoevaporation \citep{alexander09}.  A stellar
mass companion would also quicken photoevaporation since it would
increase the L$_{X}$ of the system, in addition to lowering or even
eliminating the accretion onto the star.

Another feature of EUV photoevaporation models is the uncertainty in the EUV ionizing flux since EUV emission
cannot be directly observed.
If the creation of [\ion{Ne}{3}] in SZ~Cha is being dominated by EUV radiation, this provides the most direct measurement to date of the EUV emission of TTS.  
The [Ne III] line luminosity in SZ~Cha
implies L$_{EUV}$ of $\sim$10$^{32}$--10$^{33}$ ergs~s$^{-1}$
(0.03--0.26~L$_{\sun}$)
according to the predictions of the HG09 model (based
on L$_{EUV}$$\sim$$\nu^{-1}$).
These values can be compared to SZ~Cha's L$_{acc}$ \citep[0.06
L$_{\sun}$;][]{espaillat11} and  L$_{FUV}$
\citep[0.002~L$_{\sun}$;][]{ingleby11}.
L$_{EUV}$ is of the same order as L$_{acc}$, which was determined
using U-band data with an intrinsic uncertainty of a factor
of 2--3 \citep{calvet04}, so
the EUV could in principle be powered by accretion.
The expected L$_{EUV}$ is more than a factor of 10 higher
than the observed L$_{FUV}$.  
Using the EUV spectra of solar active regions as representative,
even with a strong Lyman continuum the ratio of EUV to FUV
luminosities is expected to be lower than this 
\citep{vernazza78}. 
However, L$_{FUV}$ is underestimated since the value is based
on ACS spectra which do not cover Ly$\alpha$, which may be
up to 90$\%$ of the total FUV luminosity \citep{herczeg04,schindhelm12}.

In conclusion, these results are an important contribution to our understanding of the effect of high energy radiation from the star on the disk.  Unfortunately, the [\ion{Ne}{3}] line is not
observable from the ground, but {\it JWST} will allow us to fully
exploit the potential of this line.

\section{Summary} \label{sed:sum}

We reported new Ne forbidden line detections in high-resolution {\it Spitzer} IRS
spectra of disks in IC~348, NGC~2068, and Chamaeleon.
We use these
results, in conjunction with other reports in the literature, to arrive
at the following results and conclusions:
\\
1. Previous
to this work the only detections of [\ion{Ne}{3}] in disks were reported around SZ~102 and
TW~Hya.   Here we reported the detection of [\ion{Ne}{3}] in CS~Cha, SZ~Cha, and T~54.   
CS~Cha and T~54 have [\ion{Ne}{3}]$/$[\ion{Ne}{2}]$<$1 which is consistent
with previous [\ion{Ne}{3}]-to-[\ion{Ne}{2}] ratios.  Such small ratios can be due to Ne forbidden
line production by either soft EUV radiation or X-ray radiation.
SZ~Cha is the first object observed to have a  [\ion{Ne}{3}]-to-[\ion{Ne}{2}] ratio of
about 1.  Given that X-rays are more efficient at
producing \ion{Ne}{2} relative to \ion{Ne}{3} (GNI07, HG09), [\ion{Ne}{3}]$/$[\ion{Ne}{2}]$\sim$1
  indicates that EUV emission dominates the creation of Ne ions in SZ~Cha.   %\\
\\
2. 
In order for EUV radiation to dominate the ionization of Ne, the EUV luminosity of the object has
to be two times greater than the X-ray luminosity
(HG09).  While we do not have a direct measurement of the EUV
emission of SZ~Cha, we find that the [\ion{Ne}{3}]-to-[\ion{Ne}{2}] ratio increases as
the X-ray luminosity (normalized by L$_{bol}$) decreases. This is
consistent with the EUV luminosity being greater than the X-ray
luminosity in objects with high [\ion{Ne}{3}]-to-[\ion{Ne}{2}] line ratios.
\\
3. 
Lastly, using the HG09 models we estimate that the EUV luminosity of
SZ~Cha is $\sim$10$^{32}$--10$^{33}$ ergs~s$^{-1}$
(0.03--0.26~L$_{\sun}$).

 \acknowledgments{
We thank U. Gorti, D. Hollenbach, and I. Pascucci for comments on the manuscript and J. Raymond for discussions.  We thank the referee for useful comments.   C.~E.~was supported by the NSF under Award No. 0901947 and a Sagan Exoplanet Fellowship from the National Aeronautics and Space Administration and administered by the NASA Exoplanet Science Institute (NExScI).  
 N.~C.~acknowledges support from NASA Origins Grant NNX08AH94G. 
Support was also provided by NASA through award JPL~1309768 and Chandra award GOX-9029X. 
}

\bibliographystyle{apjv2}

\begin{thebibliography}{80}
\expandafter\ifx\csname natexlab\endcsname\relax\def\natexlab#1{#1}\fi

\bibitem[{{Alexander} \& {Armitage}(2009)}]{alexander09}
{Alexander}, R.~D., \& {Armitage}, P.~J. 2009, \apj, 704, 989

\bibitem[{{Alexander} {et~al.}(2005){Alexander}, {Clarke}, \&
  {Pringle}}]{alexander05}
{Alexander}, R.~D., {Clarke}, C.~J., \& {Pringle}, J.~E. 2005, \mnras, 358, 283

\bibitem[{{Alexander} {et~al.}(2006){Alexander}, {Clarke}, \&
  {Pringle}}]{alexander06}
---. 2006, \mnras, 369, 229

\bibitem[{{Arnaud}(1996)}]{arnaud96}
{Arnaud}, K.~A. 1996, in Astronomical Society of the Pacific Conference Series,
  Vol. 101, Astronomical Data Analysis Software and Systems V, ed.
  {G.~H.~Jacoby \& J.~Barnes}, 17--+

\bibitem[{{Artymowicz} \& {Lubow}(1994)}]{artymowicz94}
{Artymowicz}, P., \& {Lubow}, S.~H. 1994, \apj, 421, 651

\bibitem[{{Baldovin-Saavedra} {et~al.}(2011){Baldovin-Saavedra}, {Audard},
  {G{\"u}del}, {Rebull}, {Padgett}, {Skinner}, {Carmona}, {Glauser}, \&
  {Fajardo-Acosta}}]{baldovin11}
{Baldovin-Saavedra}, C., {et~al.} 2011, \aap, 528, A22

\bibitem[{{Boss}(1997)}]{boss97}
{Boss}, A.~P. 1997, Science, 276, 1836

\bibitem[{{Broos} {et~al.}(2010){Broos}, {Townsley}, {Feigelson}, {Getman},
  {Bauer}, \& {Garmire}}]{broos10}
{Broos}, P.~S., {Townsley}, L.~K., {Feigelson}, E.~D., {Getman}, K.~V.,
  {Bauer}, F.~E., \& {Garmire}, G.~P. 2010, \apj, 714, 1582

\bibitem[{{Brown}(2010)}]{brown10}
{Brown}, A. 2010, in IAU Symposium, Vol. 264, IAU Symposium, ed. A.~G.
  {Kosovichev}, A.~H. {Andrei}, \& J.-P. {Roelot}, 395--400

\bibitem[{{Brown} {et~al.}(1981){Brown}, {Ferraz}, \& {Jordan}}]{brown81}
{Brown}, A., {Ferraz}, M., \& {Jordan}, C. 1981, in NASA Conference
  Publication, Vol. 2171, NASA Conference Publication, ed. R.~D. {Chapman},
  297--302

\bibitem[{{Butler} \& {Dalgarno}(1980)}]{butler80}
{Butler}, S.~E., \& {Dalgarno}, A. 1980, \apj, 241, 838

\bibitem[{{Calvet}(1997)}]{calvet97}
{Calvet}, N. 1997, in IAU Symposium, Vol. 182, Herbig-Haro Flows and the Birth
  of Stars, ed. B.~{Reipurth} \& C.~{Bertout}, 417--432

\bibitem[{{Calvet} \& {Gullbring}(1998)}]{calvet98}
{Calvet}, N., \& {Gullbring}, E. 1998, \apj, 509, 802

\bibitem[{{Calvet} {et~al.}(2004){Calvet}, {Muzerolle}, {Brice{\~n}o},
  {Hern{\'a}ndez}, {Hartmann}, {Saucedo}, \& {Gordon}}]{calvet04}
{Calvet}, N., {Muzerolle}, J., {Brice{\~n}o}, C., {Hern{\'a}ndez}, J.,
  {Hartmann}, L., {Saucedo}, J.~L., \& {Gordon}, K.~D. 2004, \aj, 128, 1294

\bibitem[{{Carr} \& {Najita}(2011)}]{carr11}
{Carr}, J.~S., \& {Najita}, J.~R. 2011, \apj, 733, 102

\bibitem[{{Clarke} {et~al.}(2001){Clarke}, {Gendrin}, \&
  {Sotomayor}}]{clarke01}
{Clarke}, C.~J., {Gendrin}, A., \& {Sotomayor}, M. 2001, \mnras, 328, 485

\bibitem[{{Cohen} {et~al.}(2003){Cohen}, {Megeath}, {Hammersley},
  {Mart{\'{\i}}n-Luis}, \& {Stauffer}}]{cohen03}
{Cohen}, M., {Megeath}, S.~T., {Hammersley}, P.~L., {Mart{\'{\i}}n-Luis}, F.,
  \& {Stauffer}, J. 2003, \aj, 125, 2645

\bibitem[{{D'Alessio} {et~al.}(2006){D'Alessio}, {Calvet}, {Hartmann},
  {Franco-Hern{\'a}ndez}, \& {Serv{\'{\i}}n}}]{dalessio06}
{D'Alessio}, P., {Calvet}, N., {Hartmann}, L., {Franco-Hern{\'a}ndez}, R., \&
  {Serv{\'{\i}}n}, H. 2006, \apj, 638, 314

\bibitem[{{Dickey} \& {Lockman}(1990)}]{dickey90}
{Dickey}, J.~M., \& {Lockman}, F.~J. 1990, \araa, 28, 215

\bibitem[{{Drake}(1999)}]{drake99}
{Drake}, J.~J. 1999, \apjs, 122, 269

\bibitem[{{Ercolano} {et~al.}(2009){Ercolano}, {Clarke}, \&
  {Drake}}]{ercolano09}
{Ercolano}, B., {Clarke}, C.~J., \& {Drake}, J.~J. 2009, \apj, 699, 1639

\bibitem[{{Ercolano} {et~al.}(2008){Ercolano}, {Drake}, {Raymond}, \&
  {Clarke}}]{ercolano08}
{Ercolano}, B., {Drake}, J.~J., {Raymond}, J.~C., \& {Clarke}, C.~C. 2008,
  \apj, 688, 398

\bibitem[{{Ercolano} \& {Owen}(2010)}]{ercolano10}
{Ercolano}, B., \& {Owen}, J.~E. 2010, \mnras, 406, 1553

\bibitem[{{Espaillat} {et~al.}(2011){Espaillat}, {Furlan}, {D'Alessio},
  {Sargent}, {Nagel}, {Calvet}, {Watson}, \& {Muzerolle}}]{espaillat11}
{Espaillat}, C., {Furlan}, E., {D'Alessio}, P., {Sargent}, B., {Nagel}, E.,
  {Calvet}, N., {Watson}, D.~M., \& {Muzerolle}, J. 2011, \apj, 728, 49

\bibitem[{{Espaillat} {et~al.}(2007){Espaillat}, {Calvet}, {D'Alessio},
  {Bergin}, {Hartmann}, {Watson}, {Furlan}, {Najita}, {Forrest}, {McClure},
  {Sargent}, {Bohac}, \& {Harrold}}]{espaillat07a}
{Espaillat}, C., {et~al.} 2007, \apjl, 664, L111

\bibitem[{{Espaillat} {et~al.}(2012){Espaillat}, {Ingleby}, {Hern{\'a}ndez},
  {Furlan}, {D'Alessio}, {Calvet}, {Andrews}, {Muzerolle}, {Qi}, \&
  {Wilner}}]{espaillat12a}
---. 2012, \apj, 747, 103

\bibitem[{{Flaccomio} {et~al.}(2009){Flaccomio}, {Stelzer}, {Sciortino},
  {Micela}, {Pillitteri}, \& {Testi}}]{flaccomio09}
{Flaccomio}, E., {Stelzer}, B., {Sciortino}, S., {Micela}, G., {Pillitteri},
  I., \& {Testi}, L. 2009, \aap, 505, 695

\bibitem[{{Flaherty} \& {Muzerolle}(2008)}]{flaherty08}
{Flaherty}, K.~M., \& {Muzerolle}, J. 2008, \aj, 135, 966

\bibitem[{{Furlan} {et~al.}(2006){Furlan}, {Hartmann}, {Calvet}, {D'Alessio},
  {Franco-Hern{\'a}ndez}, {Forrest}, {Watson}, {Uchida}, {Sargent}, {Green},
  {Keller}, \& {Herter}}]{furlan06}
{Furlan}, E., {et~al.} 2006, \apjs, 165, 568

\bibitem[{{Glassgold} {et~al.}(2007){Glassgold}, {Najita}, \&
  {Igea}}]{glassgold07}
{Glassgold}, A.~E., {Najita}, J.~R., \& {Igea}, J. 2007, \apj, 656, 515

\bibitem[{{Gorti} \& {Hollenbach}(2008)}]{gorti08}
{Gorti}, U., \& {Hollenbach}, D. 2008, \apj, 683, 287

\bibitem[{{Gorti} \& {Hollenbach}(2009)}]{gorti09a}
---. 2009, \apj, 690, 1539

\bibitem[{{G{\"u}del} {et~al.}(2007){G{\"u}del}, {Padgett}, \&
  {Dougados}}]{gudel07}
{G{\"u}del}, M., {Padgett}, D.~L., \& {Dougados}, C. 2007, in Protostars and
  Planets V, ed. B.~{Reipurth}, D.~{Jewitt}, \& K.~{Keil}, 329--344

\bibitem[{{G{\"u}del} {et~al.}(2010){G{\"u}del}, {Lahuis}, {Briggs}, {Carr},
  {Glassgold}, {Henning}, {Najita}, {van Boekel}, \& {van Dishoeck}}]{gudel10}
{G{\"u}del}, M., {et~al.} 2010, \aap, 519, A113

\bibitem[{{Guenther} {et~al.}(2007){Guenther}, {Esposito}, {Mundt}, {Covino},
  {Alcal{\'a}}, {Cusano}, \& {Stecklum}}]{guenther07}
{Guenther}, E.~W., {Esposito}, M., {Mundt}, R., {Covino}, E., {Alcal{\'a}},
  J.~M., {Cusano}, F., \& {Stecklum}, B. 2007, \aap, 467, 1147

\bibitem[{{Hartigan} {et~al.}(1995){Hartigan}, {Edwards}, \&
  {Ghandour}}]{hartigan95}
{Hartigan}, P., {Edwards}, S., \& {Ghandour}, L. 1995, \apj, 452, 736

\bibitem[{{Hartmann} {et~al.}(1998){Hartmann}, {Calvet}, {Gullbring}, \&
  {D'Alessio}}]{hartmann98}
{Hartmann}, L., {Calvet}, N., {Gullbring}, E., \& {D'Alessio}, P. 1998, \apj,
  495, 385

\bibitem[{{Herczeg} {et~al.}(2004){Herczeg}, {Wood}, {Linsky}, {Valenti}, \&
  {Johns-Krull}}]{herczeg04}
{Herczeg}, G.~J., {Wood}, B.~E., {Linsky}, J.~L., {Valenti}, J.~A., \&
  {Johns-Krull}, C.~M. 2004, \apj, 607, 369

\bibitem[{{Higdon} {et~al.}(2004){Higdon}, {Devost}, {Higdon}, {Brandl},
  {Houck}, {Hall}, {Barry}, {Charmandaris}, {Smith}, {Sloan}, \&
  {Green}}]{higdon04}
{Higdon}, S.~J.~U., {et~al.} 2004, \pasp, 116, 975

\bibitem[{{Hollenbach} \& {Gorti}(2009)}]{hollenbach09}
{Hollenbach}, D., \& {Gorti}, U. 2009, \apj, 703, 1203

\bibitem[{{Hollenbach} {et~al.}(1994){Hollenbach}, {Johnstone}, {Lizano}, \&
  {Shu}}]{hollenbach94}
{Hollenbach}, D., {Johnstone}, D., {Lizano}, S., \& {Shu}, F. 1994, \apj, 428,
  654

\bibitem[{{Houck} {et~al.}(2004){Houck}, {Roellig}, {van Cleve}, {Forrest},
  {Herter}, {Lawrence}, {Matthews}, {Reitsema}, {Soifer}, {Watson}, {Weedman},
  {Huisjen}, {Troeltzsch}, {Barry}, {Bernard-Salas}, {Blacken}, {Brandl},
  {Charmandaris}, {Devost}, {Gull}, {Hall}, {Henderson}, {Higdon}, {Pirger},
  {Schoenwald}, {Sloan}, {Uchida}, {Appleton}, {Armus}, {Burgdorf},
  {Fajardo-Acosta}, {Grillmair}, {Ingalls}, {Morris}, \& {Teplitz}}]{houck04}
{Houck}, J.~R., {et~al.} 2004, \apjs, 154, 18

\bibitem[{{Ingleby} {et~al.}(2012){Ingleby}, {Calvet}, {Herczeg}, \&
  {Brice{\~n}o}}]{ingleby12}
{Ingleby}, L., {Calvet}, N., {Herczeg}, G., \& {Brice{\~n}o}, C. 2012, \apjl,
  752, L20

\bibitem[{{Ingleby} {et~al.}(2011){Ingleby}, {Calvet}, {Hern{\'a}ndez},
  {Brice{\~n}o}, {Espaillat}, {Miller}, {Bergin}, \& {Hartmann}}]{ingleby11}
{Ingleby}, L., {Calvet}, N., {Hern{\'a}ndez}, J., {Brice{\~n}o}, C.,
  {Espaillat}, C., {Miller}, J., {Bergin}, E., \& {Hartmann}, L. 2011, \aj,
  141, 127

\bibitem[{{Kalberla} {et~al.}(2005){Kalberla}, {Burton}, {Hartmann}, {Arnal},
  {Bajaja}, {Morras}, \& {P{\"o}ppel}}]{kalberla05}
{Kalberla}, P.~M.~W., {Burton}, W.~B., {Hartmann}, D., {Arnal}, E.~M.,
  {Bajaja}, E., {Morras}, R., \& {P{\"o}ppel}, W.~G.~L. 2005, \aap, 440, 775

\bibitem[{{Kenyon} \& {Hartmann}(1995)}]{kh95}
{Kenyon}, S.~J., \& {Hartmann}, L. 1995, \apjs, 101, 117

\bibitem[{{Kim} {et~al.}(2009){Kim}, {Watson}, {Manoj}, {Furlan}, {Najita},
  {Forrest}, {Sargent}, {Espaillat}, {Calvet}, {Luhman}, {McClure}, {Green}, \&
  {Harrold}}]{kim09}
{Kim}, K.~H., {et~al.} 2009, \apj, 700, 1017

\bibitem[{{Lafreni{\`e}re} {et~al.}(2008){Lafreni{\`e}re}, {Jayawardhana},
  {Brandeker}, {Ahmic}, \& {van Kerkwijk}}]{lafreniere08}
{Lafreni{\`e}re}, D., {Jayawardhana}, R., {Brandeker}, A., {Ahmic}, M., \& {van
  Kerkwijk}, M.~H. 2008, \apj, 683, 844

\bibitem[{{Lahuis} {et~al.}(2007){Lahuis}, {van Dishoeck}, {Blake}, {Evans},
  {Kessler-Silacci}, \& {Pontoppidan}}]{lahuis07}
{Lahuis}, F., {van Dishoeck}, E.~F., {Blake}, G.~A., {Evans}, II, N.~J.,
  {Kessler-Silacci}, J.~E., \& {Pontoppidan}, K.~M. 2007, \apj, 665, 492

\bibitem[{{Lissauer} \& {Stevenson}(2007)}]{lissauer07}
{Lissauer}, J.~J., \& {Stevenson}, D.~J. 2007, in Protostars and Planets V, ed.
  B.~{Reipurth}, D.~{Jewitt}, \& K.~{Keil}, 591--606

\bibitem[{{Lubow} \& {D'Angelo}(2006)}]{lubow06}
{Lubow}, S.~H., \& {D'Angelo}, G. 2006, \apj, 641, 526

\bibitem[{{Luhman} {et~al.}(2003){Luhman}, {Stauffer}, {Muench}, {Rieke},
  {Lada}, {Bouvier}, \& {Lada}}]{luhman03}
{Luhman}, K.~L., {Stauffer}, J.~R., {Muench}, A.~A., {Rieke}, G.~H., {Lada},
  E.~A., {Bouvier}, J., \& {Lada}, C.~J. 2003, \apj, 593, 1093

\bibitem[{{Luhman} {et~al.}(2008){Luhman}, {Allen}, {Allen}, {Gutermuth},
  {Hartmann}, {Mamajek}, {Megeath}, {Myers}, \& {Fazio}}]{luhman08}
{Luhman}, K.~L., {et~al.} 2008, \apj, 675, 1375

\bibitem[{{Manoj} {et~al.}(2011){Manoj}, {Kim}, {Furlan}, {McClure}, {Luhman},
  {Watson}, {Espaillat}, {Calvet}, {Najita}, {D'Alessio}, {Adame}, {Sargent},
  {Forrest}, {Bohac}, {Green}, \& {Arnold}}]{manoj11}
{Manoj}, P., {et~al.} 2011, \apjs, 193, 11

\bibitem[{{McClure} {et~al.}(2010){McClure}, {Furlan}, {Manoj}, {Luhman},
  {Watson}, {Forrest}, {Espaillat}, {Calvet}, {D'Alessio}, {Sargent}, {Tobin},
  \& {Chiang}}]{mcclure10}
{McClure}, M.~K., {et~al.} 2010, \apjs, 188, 75

\bibitem[{{Meijerink} {et~al.}(2008){Meijerink}, {Glassgold}, \&
  {Najita}}]{meijerink08}
{Meijerink}, R., {Glassgold}, A.~E., \& {Najita}, J.~R. 2008, \apj, 676, 518

\bibitem[{{Najita} {et~al.}(2010){Najita}, {Carr}, {Strom}, {Watson},
  {Pascucci}, {Hollenbach}, {Gorti}, \& {Keller}}]{najita10}
{Najita}, J.~R., {Carr}, J.~S., {Strom}, S.~E., {Watson}, D.~M., {Pascucci},
  I., {Hollenbach}, D., {Gorti}, U., \& {Keller}, L. 2010, \apj, 712, 274

\bibitem[{{Najita} {et~al.}(2009){Najita}, {Doppmann}, {Bitner}, {Richter},
  {Lacy}, {Jaffe}, {Carr}, {Meijerink}, {Blake}, {Herczeg}, \&
  {Glassgold}}]{najita09}
{Najita}, J.~R., {et~al.} 2009, \apj, 697, 957

\bibitem[{{Neuhaeuser} {et~al.}(1995){Neuhaeuser}, {Sterzik}, {Schmitt},
  {Wichmann}, \& {Krautter}}]{neuhaeuser95}
{Neuhaeuser}, R., {Sterzik}, M.~F., {Schmitt}, J.~H.~M.~M., {Wichmann}, R., \&
  {Krautter}, J. 1995, \aap, 297, 391

\bibitem[{{Nguyen} {et~al.}(2012){Nguyen}, {Brandeker}, {van Kerkwijk}, \&
  {Jayawardhana}}]{nguyen12}
{Nguyen}, D.~C., {Brandeker}, A., {van Kerkwijk}, M.~H., \& {Jayawardhana}, R.
  2012, \apj, 745, 119

\bibitem[{{Owen} {et~al.}(2011){Owen}, {Ercolano}, \& {Clarke}}]{owen11}
{Owen}, J.~E., {Ercolano}, B., \& {Clarke}, C.~J. 2011, \mnras, 412, 13

\bibitem[{{Owen} {et~al.}(2010){Owen}, {Ercolano}, {Clarke}, \&
  {Alexander}}]{owen10}
{Owen}, J.~E., {Ercolano}, B., {Clarke}, C.~J., \& {Alexander}, R.~D. 2010,
  \mnras, 401, 1415

\bibitem[{{Pascucci} {et~al.}(2007){Pascucci}, {Hollenbach}, {Najita},
  {Muzerolle}, {Gorti}, {Herczeg}, {Hillenbrand}, {Kim}, {Carpenter}, {Meyer},
  {Mamajek}, \& {Bouwman}}]{pascucci07}
{Pascucci}, I., {et~al.} 2007, \apj, 663, 383

\bibitem[{{Pontoppidan} {et~al.}(2010){Pontoppidan}, {Salyk}, {Blake},
  {Meijerink}, {Carr}, \& {Najita}}]{pontoppidan10}
{Pontoppidan}, K.~M., {Salyk}, C., {Blake}, G.~A., {Meijerink}, R., {Carr},
  J.~S., \& {Najita}, J. 2010, \apj, 720, 887

\bibitem[{{Preibisch} {et~al.}(2005){Preibisch}, {Kim}, {Favata}, {Feigelson},
  {Flaccomio}, {Getman}, {Micela}, {Sciortino}, {Stassun}, {Stelzer}, \&
  {Zinnecker}}]{preibisch05}
{Preibisch}, T., {et~al.} 2005, \apjs, 160, 401

\bibitem[{{Ribas} {et~al.}(2005){Ribas}, {Guinan}, {G{\"u}del}, \&
  {Audard}}]{ribas05}
{Ribas}, I., {Guinan}, E.~F., {G{\"u}del}, M., \& {Audard}, M. 2005, \apj, 622,
  680

\bibitem[Rice et al.(2006)]{rice06} Rice, W.~K.~M., Armitage, 
P.~J., Wood, K., \& Lodato, G.\ 2006, \mnras, 373, 1619

\bibitem[{{Sacco} {et~al.}(2012){Sacco}, {Flaccomio}, {Pascucci}, {Lahuis},
  {Ercolano}, {Kastner}, {Micela}, {Stelzer}, \& {Sterzik}}]{sacco12}
{Sacco}, G.~G., {et~al.} 2012, \apj, 747, 142

\bibitem[{{Sanz-Forcada} {et~al.}(2002){Sanz-Forcada}, {Brickhouse}, \&
  {Dupree}}]{sanzforcada12}
{Sanz-Forcada}, J., {Brickhouse}, N.~S., \& {Dupree}, A.~K. 2002, \apj, 570,
  799

\bibitem[Schindhelm et al.(2012)]{schindhelm12} Schindhelm, E., 
France, K., Herczeg, G.~J., et al.\ 2012, \apjl, 756, L23 

\bibitem[{{Schisano} {et~al.}(2010){Schisano}, {Ercolano}, \&
  {G{\"u}del}}]{schisano10}
{Schisano}, E., {Ercolano}, B., \& {G{\"u}del}, M. 2010, \mnras, 401, 1636

\bibitem[{{Skrutskie} {et~al.}(2006){Skrutskie}, {Cutri}, {Stiening},
  {Weinberg}, {Schneider}, {Carpenter}, {Beichman}, {Capps}, {Chester},
  {Elias}, {Huchra}, {Liebert}, {Lonsdale}, {Monet}, {Price}, {Seitzer},
  {Jarrett}, {Kirkpatrick}, {Gizis}, {Howard}, {Evans}, {Fowler}, {Fullmer},
  {Hurt}, {Light}, {Kopan}, {Marsh}, {McCallon}, {Tam}, {Van Dyk}, \&
  {Wheelock}}]{skrutskie06}
{Skrutskie}, M.~F., {et~al.} 2006, \aj, 131, 1163

\bibitem[{{Smith} {et~al.}(2001){Smith}, {Brickhouse}, {Liedahl}, \&
  {Raymond}}]{smith01}
{Smith}, R.~K., {Brickhouse}, N.~S., {Liedahl}, D.~A., \& {Raymond}, J.~C.
  2001, \apjl, 556, L91

\bibitem[{{Szulagyi}(2012)}]{szulagyi12}
{Szulagyi}, e.~a. 2012, in press, ApJ

\bibitem[{{Takami} {et~al.}(2003){Takami}, {Bailey}, \&
  {Chrysostomou}}]{takami03}
{Takami}, M., {Bailey}, J., \& {Chrysostomou}, A. 2003, \aap, 397, 675

\bibitem[{{Vernazza} \& {Reeves}(1978)}]{vernazza78}
{Vernazza}, J.~E., \& {Reeves}, E.~M. 1978, \apjs, 37, 485

\bibitem[{{Vuong} {et~al.}(2003){Vuong}, {Montmerle}, {Grosso}, {Feigelson},
  {Verstraete}, \& {Ozawa}}]{vuong03}
{Vuong}, M.~H., {Montmerle}, T., {Grosso}, N., {Feigelson}, E.~D.,
  {Verstraete}, L., \& {Ozawa}, H. 2003, \aap, 408, 581

\bibitem[{{Watson} {et~al.}(2009){Watson}, {Leisenring}, {Furlan}, {Bohac},
  {Sargent}, {Forrest}, {Calvet}, {Hartmann}, {Nordhaus}, {Green}, {Kim},
  {Sloan}, {Chen}, {Keller}, {d'Alessio}, {Najita}, {Uchida}, \&
  {Houck}}]{watson09}
{Watson}, D.~M., {et~al.} 2009, \apjs, 180, 84

\bibitem[{{Werner} {et~al.}(2004){Werner}, {Roellig}, {Low}, {Rieke}, {Rieke},
  {Hoffmann}, {Young}, {Houck}, {Brandl}, {Fazio}, {Hora}, {Gehrz}, {Helou},
  {Soifer}, {Stauffer}, {Keene}, {Eisenhardt}, {Gallagher}, {Gautier}, {Irace},
  {Lawrence}, {Simmons}, {Van Cleve}, {Jura}, {Wright}, \&
  {Cruikshank}}]{werner04}
{Werner}, M.~W., {et~al.} 2004, \apjs, 154, 1

\bibitem[{{White} {et~al.}(2000){White}, {Giommi}, \& {Angelini}}]{white00}
{White}, N.~E., {Giommi}, P., \& {Angelini}, L. 2000, VizieR Online Data
  Catalog, 9031, 0

\bibitem[{{Wilms} {et~al.}(2000){Wilms}, {Allen}, \& {McCray}}]{wilms00}
{Wilms}, J., {Allen}, A., \& {McCray}, R. 2000, \apj, 542, 914

\bibitem[{{Zhu} {et~al.}(2011){Zhu}, {Nelson}, {Hartmann}, {Espaillat}, \&
  {Calvet}}]{zhu11}
{Zhu}, Z., {Nelson}, R.~P., {Hartmann}, L., {Espaillat}, C., \& {Calvet}, N.
  2011, \apj, 729, 47

\end{thebibliography}

\appendix

\section{Sample Descriptions}\label{appendixsample}

\begin{figure}
\epsscale{1}
\plotone{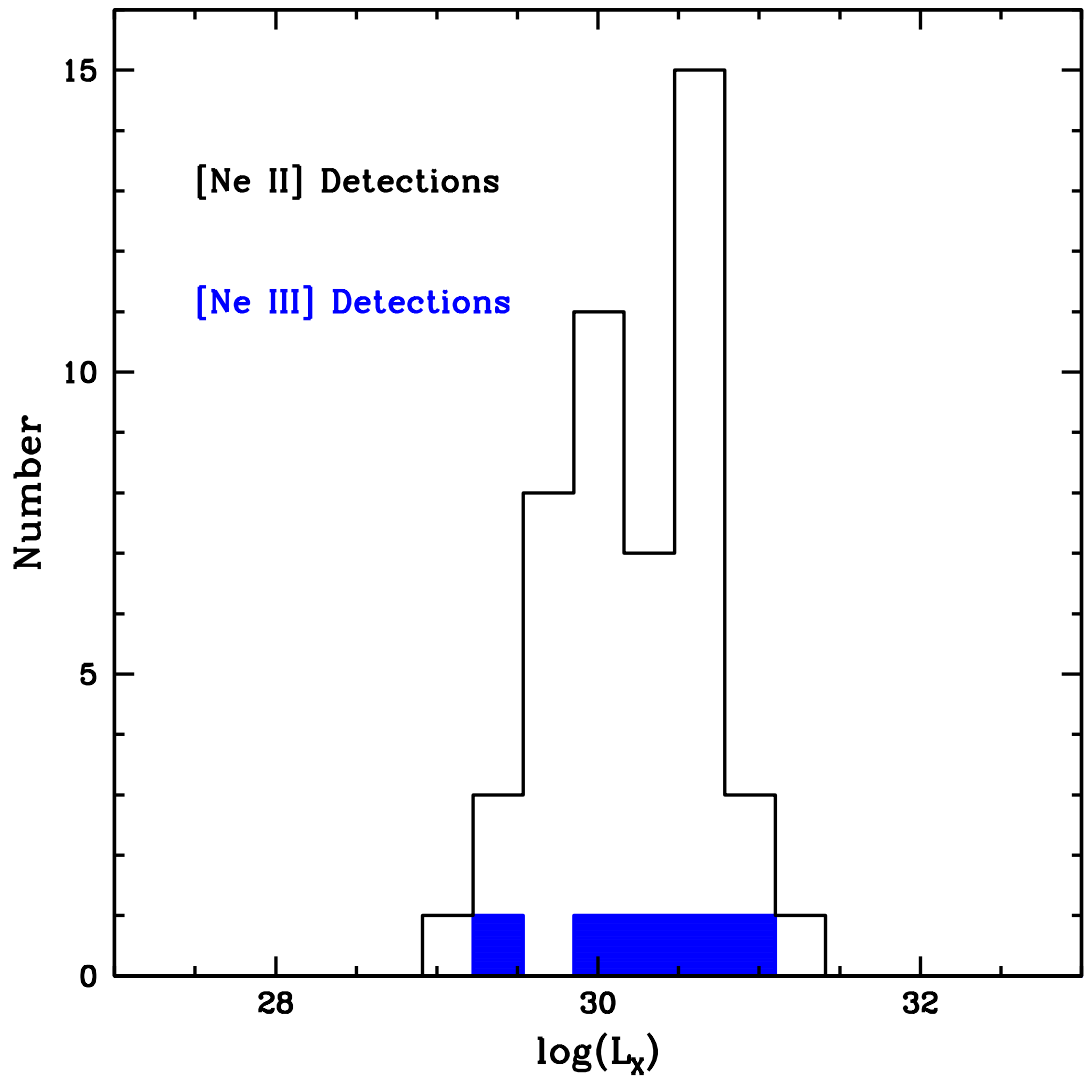}
\caption[]{
X-ray luminosity of disks with [\ion{Ne}{2}] (black) and disks with [\ion{Ne}{3}] (blue).  

}
\label{figxray}
\end{figure}

\begin{figure}
\epsscale{1}
\plotone{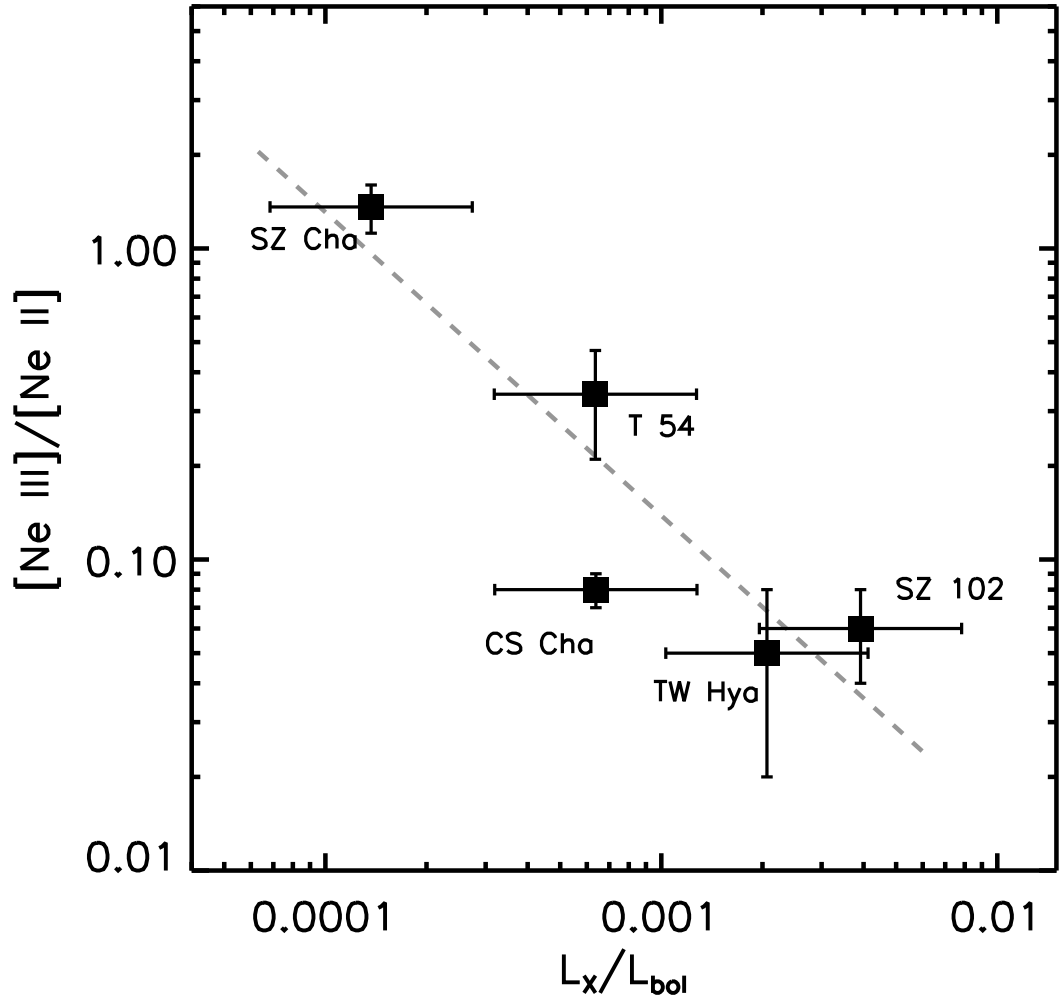}
\caption[]{
Dependence of the [\ion{Ne}{3}]-to-[\ion{Ne}{2}] ratio on L$_{X}$$/$L$_{bol}$.  
Given the typical range in X-ray variability, we assume that each L$_{X}$ 
is uncertain within a factor of 2 of the measured value,
and that this dominates
the errors for L$_{X}$$/$L$_{bol}$, as has been assumed in previous works \citep{gudel10}.
The regression line for the sample is the broken gray line.

}
\label{figreln}
\end{figure}

For comparison with the results from our sample (described in
Section~\ref{sampleselection}), we compiled additional reports of
[\ion{Ne}{2}] line emission from the literature. We limited ourselves to studies of Class~II disks
using {\it Spitzer} IRS spectra.  We note that \citet{flaccomio09}
report [\ion{Ne}{2}] detections in the disks of WL~10, IRS~45, and IRS~47. 
However, we do not include these detections in this work since they are
located in areas of high extinction  \citep[A$_V$$>$24;][]{mcclure10}
and so their reported line fluxes are less certain.  We also
note that \citet{gudel10} report a [\ion{Ne}{2}] detection in SZ~50, but
we do not see this line in our analysis of the spectrum and so do not list
SZ~50 as having a [\ion{Ne}{2}] detection here.
The literature sample used in Figure~\ref{figxray} consists of the objects listed in Table~\ref{sample1}.
These objects have a reported [\ion{Ne}{2}] detection as well as
known X-ray luminosities from Section~\ref{xprop} or from the
literature.   

\clearpage

\begin{deluxetable}{cccc}
\tabletypesize{\scriptsize}
\tablewidth{0pt}
\tablecaption{SH GO Sample\label{tab:targetlog}}
\startdata
\hline
\hline
Object &  Region & RA & DEC \\
\hline
CS~Cha 	& Cha 			& 11h02m25s & --77d33m36s\\  
FM~177 &  NGC 2068 	& 05h45m42s & --00d12m05s   \\  
FM~281 &	 NGC 2068 	& 05h45m53s & --00d13m25s    \\  
FM~515 &	 NGC 2068 	& 05h46m12s  & +00d32m26s   \\ 
FM~581 &	 NGC 2068 	& 05h46m19s & --00d05m38s  \\
FM~618 &	  NGC 2068 	& 05h46m23s & --00d08m53s    \\ 
LRLL~2 &	   IC 348			& 03h44m35s & +32d10m04s   \\ 
LRLL~6 &	   IC 348			& 03h44m37s & +32d06m45s  \\ 
LRLL~21	&	  IC 348			& 03h44m56s & +32d09m15s \\ 
LRLL~31 	&	  IC 348			& 03h44m18s & +32d04m57s  \\ 
LRLL~37 	& IC 348			& 03h44m38s & +32d03m29s  \\  
LRLL~55 	&	 IC 348			&  03h44m31s & +32d00m14s \\  
LRLL~67	&	IC 348			& 03h43m45s & +32d08m17s   \\ 
LRLL~68 	&	IC 348			& 03h44m29s & +31d59m54s \\  
LRLL~72 		&	IC 348			& 03h44m23s & +32d01m53s  \\  
LRLL~133 		& IC 348			& 03h44m42s & +32d12m02s  \\
SZ~Cha 	& Cha 			& 10h58m17s & --77d17m17s\\  
T~35 		& Cha 			& 11h08m39s & --77d16m04s \\
T~54 		& Cha 			& 11h12m43s & --77d22m23s    

\enddata
\tablecomments{
Target ID's are taken from
\citet{flaherty08} and  \citet{luhman03} for targets in NGC~2068 and
IC~348, respectively.  Alternate names for the Chamaelon targets are T~11 (CS~Cha), T~6
(SZ~Cha), CHX~22 (T~54), FL~Cha, SZ~27, and HM~32 (T~35).
}
\end{deluxetable}

\begin{deluxetable}{lllllllllll}
\tabletypesize{\scriptsize}
\tablewidth{0pt}
\tablecaption{Source Properties \label{tab:prop}}
\startdata
\hline
\hline
Object & Disk & A$_V$ & Spectral  & T$_{*}$ &  L$_{*}$          & M$_{*}$         & R$_{*}$          &  $\mdot$                       		             & L$_X$ & Ref. \\
             &  Type &             & Type        & (K)           & (M$_{\sun}$) & (M$_{\sun}$) & (R$_{\sun}$) &  (10$^{-8}$ M$_{\sun}$ yr$^{-1}$) & (L$_{\sun}$) &   \\
 ~~~~(1)    & (2)    & (3)        &(4)                 &(5)           & (6)                       & (7)                    &(8)                                                            &(9)                      &   (10) & (11) \\                 
\hline
CS~Cha &  TD   & 0.8 & K6      	& 4205 & 1.5 & 0.9     & 2.3 & 1.2        & 8.3$\times$	10$^{-4}$     		& 1, 2 \\
FM~177 	& TD & 1.6 & K4 	& 4590 & 1.0 & 1.2 	& 1.5 & 0.004 	& 3.16$\times$10$^{-4}$			& 3, 4   \\  
FM~281 	& TD & 2.0 & M1 	& 3720 & 0.4 & 0.5 	& 1.6 & 0.002 	& 2.62$\times$10$^{-4}$			& 3, 4  \\  
FM~515 	& PTD & 1.6 & K2 	& 4900 & 2.5 & 1.5 	& 2.2 & 3.10 & ...							& 3, 4   \\ 
FM~581 	& PTD & 4.1 & K4 	& 4590 & 4.1 & 1.6 	& 3.1 & 2.57 & 1.45$\times$10$^{-3}$ 			& 3, 4   \\
FM~618 	& FD & 2.9 & K1 	& 5080 & 2.2 & 1.5 	& 1.9	 & 1.21  & 8.90$\times$10$^{-4}$ 			& 3, 4  \\ 
LRLL~2	& FD & 3.8 & A2 	& 8970 & 57.1 & 2.8 & 3.1  		& ...	 & 1.81$\times$10$^{-3}$		&  3, 4  \\ 
LRLL~6	& FD & 3.9 & G3 	& 5830 & 16.6 & 2.4 	& 4.0 	 	& ...	& 1.72$\times$10$^{-3}$		& 3, 4  \\ 
LRLL~21	& PTD & 4.7 & K0 	& 5250 & 3.8 & 1.6 	& 2.4 & 0.20 		& 7.96$\times$10$^{-4}$ 		& 3, 4   \\ 
LRLL~31	& PTD & 8.6 & G6 	& 5700 & 5.0 & 1.6 	& 2.3 & 1.4 		& 2.62$\times$10$^{-1}$		& 3, 4  \\ 
LRLL~37 	& PTD & 2.8 & K6 	& 4205 & 1.3 & 0.9 	& 2.2 & 0.13 		& 1.86$\times$10$^{-4}$		& 3, 4    \\  
LRLL~55 	& FD & 8.5 & M0.5 	& 3850 & 1.0 & 0.6 	& 2.2 & ...				&  2.04$\times$10$^{-4}$	& 3, 4    \\  
LRLL~67	& TD & 2.0 & M0.75 	& 3720 & 0.5 & 0.5 	& 1.8 & 0.01 		& 1.31$\times$10$^{-3}$		& 3, 4  \\ 
LRLL~68 	& FD & 2.1 & M3.5	 &3470 & 0.5 & 0.3 	& 2.0 & 0.04 		& $<$6.03$\times$10$^{-6}$	& 3, 4   \\  
LRLL~72 	& TD	& 3.0 & M2.5 	& 3580 & 0.7 & 0.4 	& 2.1 & $<$0.0003	 & 3.85$\times$10$^{-5}$		& 3, 4  \\  
LRLL~133 & TD & 3.6 & M5 	& 3240 & 0.2 & 0.2	& 1.5 & $<$0.8		& 3.18$\times$10$^{-5}$		& 3, 4  \\  
SZ~Cha  & PTD & 1.9 & K0 	& 5250 & 1.9 & 1.4	 & 1.7 & 0.24 	 & 2.6$\times$	10$^{-4}$			& 1, 2  \\  
T~35 	& PTD & 3.5 & M0 	& 3850 & 0.4 & 0.6 	& 1.5 & 0.12 		& 3.3$\times$10$^{-5}$		& 1, 2  \\
T~54 	& TD & 1.8 & G8  	& 5520 & 3.3 & 1.5 	& 2.0 & ... 		& 2.1$\times$10$^{-3}$			& 5, 2
\enddata
\tablecomments{
Col. (1): Name of target.
Col. (2): We label objects as transitional disks (TD), pre-transitional disks (PTD), and full disks (FD). 
Col. (3): Visual extinction.
Col. (4): Spectral type.
Col. (5): Stellar temperature.
Col. (6): Stellar luminosity.
Col. (7): Stellar mass.
Col. (8): Stellar radius.
Col. (9): Mass accretion rate.
Col. (10): X-ray luminosity.
Col. (11): Literature references for the values listed in Cols. (1)--(10).
Disk Type, A$_{V}$, L$_{*}$, M$_{*}$, R$_{*}$, and $\mdot$ are the same as those adopted in 
[1] \citet{espaillat11},
[3] \citet{espaillat12a},
and
[5] Espaillat et al, (in prep).
L$_{X}$ is from [2] \citet{ingleby11}, and
[4] this work.
We assume a distance of 160~pc for Chamaeleon \citep{luhman03}, 315~pc for IC~348  \citep{luhman08}, and 400~pc
for NGC~2068 \citep{flaherty08}.
} 
\end{deluxetable}

\begin{deluxetable}{cccc}
\tabletypesize{\scriptsize}
\tablewidth{0pt}
\tablecaption{Log of {\it Spitzer} IRS Observations \label{tab:irslog}}
\startdata
\hline
\hline
Object & AOR ID  & Date & Exposure (s) \\
\hline
CS~Cha 		&  18021632 & 2006-08-02 & 1500\\  
FM~177 		& 22852864 & 2008-11-18 & 6000 \\  
FM~281		& 22853376 & 2008-11-20 & 6000 \\  
FM~515  		& 22853888 & 2008-04-23 & 3000 \\ 
FM~581 		& 22854400 & 2008-04-23 & 2400 \\ 
FM~618  		& 22854912 & 2008-04-23 & 1500\\
LRLL~2		& 22847744 & 2007-10-13 & 120 \\ 
LRLL~6		& 22848256 & 2008-10-17 & 480 \\ 
LRLL~21		& 22848768 & 2008-10-14& 900 \\ 
LRLL~31		& 22849280 & 2008-10-17 & 480 \\ 
LRLL~37 		& 22849792 & 2008-10-14 & 900 \\  
LRLL~55 		& 22850304 & 2008-10-14 & 2400\\  
LRLL~67		& 22850816 & 2008-10-17 & 3000\\ 
LRLL~68 		& 22851328 & 2008-10-17 & 3000 \\  
LRLL~72 		& 22851840 & 2008-10-14 & 3000 \\  
LRLL~133 	& 22852352 & 2008-10-13 & 3000 \\  
SZ~Cha 		&  22846208 & 2008-08-17 & 120\\  
T~35 		& 22847232 & 2008-10-12 & 3000 \\
T~54 		&  22846720 & 2008-10-12 & 1800  
\enddata
\end{deluxetable}

\begin{deluxetable}{cccc}
\tabletypesize{\scriptsize}
\tablewidth{0pt}
\tablecaption{Log of {\it Chandra} ACIS Observations \label{tab:chandralog}}
\startdata
\hline
\hline
Target & ObsID & Date & Exposure (ks) \\
\hline
IC 348 				& 8584 & 2008-03-15 & 50 \\  
NGC 2068 		& 8585 & 2008-11-28 & 29 \\
NGC 2068		& 10763 & 2008-11-27 & 20 
\enddata
\end{deluxetable}

\clearpage
\begin{deluxetable}{crccccc}

\centering \tabletypesize{\scriptsize} \tablewidth{0pt}

\tablecaption{X-ray Spectroscopy for IC~348 and NGC~2068:  Thermal Plasma Fits
\label{tbl:thermal_spectroscopy}}

\startdata
\hline
\hline
%\tablehead{

\multicolumn{2}{c}{Source\tablenotemark{a}} &
\multicolumn{3}{c}{Spectral Fit\tablenotemark{b}} &      % 0 abundances
\multicolumn{2}{c}{X-ray Luminosities\tablenotemark{c}} \\ 

\multicolumn{2}{c}{\hrulefill} &
\multicolumn{3}{c}{\hrulefill} &       % 0 abundances
\multicolumn{2}{c}{\hrulefill} \\

\colhead{Object} & \colhead{$C_{net}$} & 
\colhead{$N_H$} & \colhead{$kT$} &   \colhead{$\chi^2$} &
\colhead{$L_{X}$} & \colhead{$L_{X,corr}$} \\

\colhead{} & \colhead{} & 
\colhead{(10$^{22}$ cm$^{-2}$)} & \colhead{(keV)} & \colhead{} &
\colhead{(10$^{30}$ ergs~s$^{-1}$)} & \colhead{(10$^{30}$ ergs~s$^{-1}$)} \\

 ~~~~(1)    & (2)    & (3)        &(4)                 &(5)           & (6)                       & (7)            \\   
\hline
%\startdata
FM~177  		&   288.0 		&  0.78  	&$0.34,2.02$$^{d}$  	& $0.28$       	& 1.02	& 4.47    \\
FM~281  	& 71.5       		& 1.07    	&$0.68$                  	& 0.77          	& 0.19   	& 1.00   	  \\
FM~581 	&   537.5   	&   0.32    	&$2.63$                	& $ 1.37$     	& 4.20  	&  5.56   	   \\
FM~618 &   186.8  		& 0.93 	&$0.36,4.82$$^{d}$          & $0.42$  		& 0.69	& 3.41 \\
LRLL~2   	&   1097.7 	&    0.26  	&$1.85$               	& $1.53 $    	& 2.57   	& 3.45   	  \\
LRLL~6   	&   1630.5 	&    0.55 	&$1.59$                	& $0.93$     	& 3.77   	& 6.58 	  \\
LRLL~21 	&   675.7 		&   0.56 	&$3.22$                  	& $ 0.95$     	& 2.12  	& 3.05  	 \\
LRLL~31 	&   184.8 		&   2.20     &$1.49$                  	&  $1.04$      	& 0.60  	& 2.06    	 \\
LRLL~37 	& 109.9   		&  0.23 	&$4.75$                      	& 0.35           	& 0.60    	&  0.71  	  \\
LRLL~55 	&   157.0 		&   0.65 	&$5.77$                  	&   $ 0.69$  	& 0.58  	& 0.78    	 \\
LRLL~67 	&   194.3 		&   0.74 	&$0.20,0.83$$^{d}$        &  $1.00$     	& 0.42	&  5.03  \\
LRLL~68 	&   ... 		&   ... 	& ...        &  ...    	& ...	&  $<$0.02$^{e}$  \\
LRLL~72 	&  63.4  		&  0.04  	&$1.43$                    	& 1.36           	& 0.14 	& 0.15        \\
LRLL~133 	&  30.3  		& 0.93  	&$ 2.46$                    	& 0.73            	& 0.06  	& 0.12     
\enddata

\tablenotetext{a}{ 
Columns\ (1)--(2) list the target names and net counts.  We note that FM~515 did not lie in the FOV of our observations and is not included here.
}

\tablenotetext{b}{
Cols.\ (3) and (4) present the best-fit values for the extinction column density and plasma temperature parameters.
Col.\ (5) presents the $\chi^{2}$ of the model fit. 
}
\tablenotetext{c}{ X-ray luminosities derived from the model spectrum are presented in cols.\ (6) and (7) calculated over the band 0.5--8 keV.  
Absorption-corrected luminosities are subscripted with $corr$. 
}
\tablenotetext{d}{ Here a two-temperature model was used and both temperatures are listed.} 

\tablenotetext{e}{
Object LRLL~68 was not detected and here we list an upper limit.  See Section~\ref{xprop} for more details.
} 

\end{deluxetable}

\begin{deluxetable}{ccccccc}
\tabletypesize{\scriptsize}
\tablewidth{0pt}
\tablecaption{Neon Line Fluxes \label{tab:neon}}
\startdata
\hline
\hline

\multicolumn{1}{c}{Object} &
%\multicolumn{2}{c}{X-ray Luminosities} &
\multicolumn{3}{c}{[Ne II]} &
\multicolumn{3}{c}{[Ne III]}  \\

\multicolumn{1}{c}{} &
\multicolumn{3}{c}{\hrulefill} &
\multicolumn{3}{c}{\hrulefill} \\

\multicolumn{1}{c}{} &
\multicolumn{1}{c}{Flux} &
\multicolumn{1}{c}{Error} &
\multicolumn{1}{c}{SNR}  &
\multicolumn{1}{c}{Flux} &
\multicolumn{1}{c}{Error} &
\multicolumn{1}{c}{SNR}  \\

\multicolumn{1}{c}{} &
\multicolumn{2}{c}{(10$^{-15}$ erg~cm$^{-2}$~s$^{-1}$)} &
\multicolumn{1}{c}{} &
\multicolumn{2}{c}{(10$^{-15}$ erg~cm$^{-2}$~s$^{-1}$)} &
\multicolumn{1}{c}{}

\\
\hline
CS~Cha 	         & 36.3 		& 0.65 		& 105   	& 3.07 	 & 0.48 & 12 \\  
FM~177  		& $<$1.56 	& $. . .$ 		& $. . .$     	& $<$1.24 & $. . .$ & $. . .$\\ 
FM~281  		& $<$1.89		& $. . .$		& $. . .$ 	& $<$1.40 & $. . .$    & $. . .$ \\ 
FM~515  		& $<$1.41 	& $. . .$ 		& $. . .$     	& $<$1.24 & $. . .$ & $. . .$\\ 
FM~581 		& 2.05 		& 0.73  		& 5     	& $<$1.29 & $. . .$ & $. . .$\\ 
FM~618  		& 1.82 		& 0.73 		& 4    	& $<$1.37 & $. . .$ & $. . .$\\
LRLL~2		& 7.95 		& 1.13 		& 12 		& $<$5.52 & $. . .$ & $. . .$ \\ 
LRLL~6		& $<$5.20 	& $. . .$  		& $. . .$     	& $<$2.72 & $. . .$ & $. . .$\\ 
LRLL~21		&$<$5.12  	& $. . .$  		&  $. . .$  	& $<$2.93 & $. . .$ & $. . .$ \\ 
LRLL~31		& $<$6.22  	& $. . .$ 		& $. . .$  	& $<$3.46 & $. . .$ & $. . .$ \\ 
LRLL~37 		& $<$2.00		& $. . .$ 	 	& $. . .$   	& $<$2.18 & $. . .$  & $. . .$  \\  
LRLL~55 		& 2.16 		& 1.58  		&  6  		& $<$2.91 & $. . .$ & $. . .$ \\  
LRLL~67		& 2.83 		& 0.91 		& 8  		& $<$1.72 & $. . .$  & $. . .$ \\ 
LRLL~68 		& $<$2.10 	& $. . .$  		& $. . .$	& $<$1.25 & $. . .$ & $. . .$\\  
LRLL~72 		& $<$1.83 	& $. . .$  		& $. . .$    	& $<$1.74 & $. . .$ & $. . .$\\  
LRLL~133 	& $<$2.91 	& $. . .$  		& $. . .$    	& $<$1.90 & $. . .$ & $. . .$ \\  
SZ~Cha 	         & 16.2 		& 1.96 		& 14  	& 22.0 	& 2.72 & 11 \\  
T~35 		& 6.53 		& 0.69 		& 18   	& $<$1.53 & $. . .$ & $. . .$\\
T~54 		& 6.15 		& 1.20 		& 7 		& 2.11 	& 0.66 & 6
\enddata
\end{deluxetable}

\begin{deluxetable}{lccc}
\tabletypesize{\scriptsize}
\tablewidth{0pt}
\tablecaption{Reported [\ion{Ne}{2}] Detections in Disks \& X-ray Luminosities \label{sample1}}
\startdata
\hline
\hline
Target	&			[\ion{Ne}{2}]	&	L$_{X}$	& L$_{X}$\\
	&			Detection	&		(ergs~s$^{-1}$) 	& Reference   \\
\hline
AA Tau				& CN11* 	& 1.0$\times$10$^{30}$ &	G10 \\
BP Tau				& G10 	& 1.4$\times$10$^{30}$ & 	G10 \\
CoKu Tau$/$3			& G10 	& 5.7$\times$10$^{30}$ & 	G10 \\
CS Cha				& E12*	& 3.8$\times$10$^{30}$	 &   I11 \\
DG Tau				& G10 	& 5.5$\times$10$^{29}$	 &	G10 \\
DK Tau				& G10 	& 8.4$\times$10$^{29}$ & 	G10 \\
DM Tau				& E07 	& 2.0$\times$10$^{30}$ &	G10 \\
DO Tau				& CN11* 	& 2.4$\times$10$^{29}$ &	G10 \\
DoAr 25				& G10 	& 2.8$\times$10$^{30}$ &	G10 \\
EC 74				& G10 	& 4.3$\times$10$^{30}$ & 	G10	\\
EC 82				& G10 	& 9.2$\times$10$^{28}$ &	G10 \\
EC 92				& G10 	& 9.5$\times$10$^{30}$ &  	G10 \\
FM 581				& E12	& 5.6$\times$10$^{30}$ &	E12 \\
FM 618				& E12	& 3.4$\times$10$^{30}$ &	E12	\\	
FS Tau A				& B11 	& 3.2$\times$10$^{30}$ &	B11	\\
GI Tau				& G10 	& 6.7$\times$10$^{29}$ &	G10 \\
GK Tau				& G10 	& 1.2$\times$10$^{30}$ &	G10 \\
GM Aur				& G10 	& 1.6$\times$10$^{30}$ & 	G10 \\
GQ Lup				& E12 	& 7.4$\times$10$^{29}$ &	G10 \\
Haro 1-4				& G10 	& 4.1$\times$10$^{29}$ &	W00	\\
Haro 1-16     			& G10 	& 1.3$\times$10$^{30}$ &	G10 \\
Haro 1-17				& G10 	& 5.8$\times$10$^{29}$ &	G10 \\
IM Lup				& G10 	& 3.2$\times$10$^{30}$ &	G10 \\
IP Tau				& G10 	& 6.1$\times$10$^{29}$ &	I11 \\
IQ Tau				& G10 	& 3.2$\times$10$^{29}$ &	G10 \\
IRAS 08267-3336		& G10 	& 1.8$\times$10$^{31}$ &	G10 \\
IRS 51				& G10 	& 3.4$\times$10$^{30}$ &	G10	\\
IRS 60				& G10 	& 2.6$\times$10$^{29}$ &	G10	\\
LkH$\alpha$ 270		& G10 	& 1.1$\times$10$^{31}$ &	G10 \\
LRLL 2				& E12 	& 3.5$\times$10$^{30}$ &	E12 \\
LRLL 55				& E12	& 0.8$\times$10$^{30}$ &	E12	\\
LRLL 67				& E12	& 5.0$\times$10$^{30}$ &	E12 \\
PZ99 J161411			& G10 	& 3.5$\times$10$^{30}$ &	P07 \\
Rox 42C 				& G10 	& 4.5$\times$10$^{30}$ &	G10	\\
RU Lup				& G10* 	& 1.0$\times$10$^{30}$ &	G10  \\
RW Aur				& CN11* 	& 1.6$\times$10$^{30}$	 &	B11 \\
RXJ1111.7-7620		& G10 	& 3.6$\times$10$^{30}$ &	P07 \\
RXJ1842.9-3542		& G10	& 2.3$\times$10$^{30}$ &	P07 \\
RXJ1852.3-3700     		& G10	& 3.7$\times$10$^{30}$ &	P07 \\
SY Cha				& G10 	& 6.9$\times$10$^{29}$ &	G10 \\
SZ Cha				& E12	& 9.9$\times$10$^{29}$ &	I11	\\
Sz 102				& L07	& 1.8$\times$10$^{29}$ & 	G10	\\
T Cha				& G10 	& 1.1$\times$10$^{30}$ & 	G10 \\
TW Hya				& N10	& 2.1$\times$10$^{30}$ &	G10\\
T 54					& E12	& 8.0$\times$10$^{30}$ &	I11\\
UY Aur				& CN11* 	& 4.0$\times$10$^{29}$	 &	G10 \\
VW Cha				& G10 	& 2.5$\times$10$^{30}$ &	I11	\\
VZ Cha				& G10 	& 5.3$\times$10$^{29}$ &	G10	\\
V773 Tau				& B11 	& 9.5$\times$10$^{30}$ &	B11 \\
V836 Tau				& G10 	& 1.7$\times$10$^{30}$ &	G10 \\
V853 Oph				& G10 	& 3.1$\times$10$^{30}$ &	G10 \\
V4046 Sgr			& S12 	& 1.2$\times$10$^{30}$ &	S12 \\
WX Cha				& S12 	& 4.6$\times$10$^{30}$ &	G10 \\	
XX Cha				& G10 	& 1.1$\times$10$^{30}$ &	G10 
\enddata
\tablecomments{
For each target, the reference for the [\ion{Ne}{2}] line detection is indicated
and an asterisk indicates the source has a known jet.  X-ray luminosities
and relevant references are also listed.
References are G10: Gudel et al. 2010; B11: Baldovin et al. 2011; CN11: Carr \& Najita 2011; 
E12: this work; S12: Sacco et al. 2012; L07: Lahuis et al. 2007; N10: Najita et al. 2010; 
P07: Pascucci et al. 2007; I11: Ingleby et al. 2011; W00: White et al. 2000.  
}
\end{deluxetable}

\clearpage

\clearpage

\clearpage

\clearpage

\end{document}